%% file: main.tex
\documentclass[preprint,12pt,authoryear]{elsarticle}
\usepackage{amssymb}
\usepackage{amsthm}
\usepackage{amsmath}
\usepackage{gensymb}
\usepackage{graphicx}
\usepackage{tabularx}
\usepackage{physics}
\usepackage{commath}
\usepackage{tikz}
\usetikzlibrary{shapes,patterns,spy,external}
\usepackage{pgfplots}
\usepackage[outline]{contour}
\usepackage{color,soul}
\usetikzlibrary{decorations.markings}
\usepackage{multirow}
\usepackage{comment}
\usepackage{transparent}
\newcommand{\ubar}[1]{\text{\b{$#1$}}}
\graphicspath{{figures/}}

\usepackage{soul}
\soulregister\cite7
\soulregister\ref7

\journal{arXiv}
\begin{document}
\begin{frontmatter}
	\title{Topology optimization for 3D thin-walled structures with adaptive meshing}
	\author[1,2]{Yuqing Zhou}
	\author[2,3]{Tsuyoshi Nomura}
	\author[2]{Ercan M. Dede}
	\author[1]{Kazuhiro Saitou}
	\address[1]{Department of Mechanical Engineering, University of Michigan, 2350 Hayward, Ann Arbor, Michigan, 48109, USA}
	\address[2]{Toyota Research Institute of North America, 1555 Woodridge Avenue, Ann Arbor, Michigan, 48105, USA}
	\address[3]{Toyota Central R \& D Labs., Inc., 41-1 Yokomichi, Nagakute 480-1192, Japan}
	\begin{abstract}
		This paper presents a density-based topology optimization method for designing 3D thin-walled structures with adaptive meshing. Uniform wall thickness is achieved by simultaneously constraining the minimum and maximum feature sizes using Helmholtz partial differential equations (PDE).  The PDE-based constraints do not require information about neighbor cells and therefore can readily be integrated with an adaptive meshing scheme.  This effectively enables the 3D topology optimization of thin-walled structures with a desktop PC, by significantly reducing computation in large void regions that appear during optimization. The uniform feature size constraint, when applied to 3D structures, can produce thin-walled geometries with branches and holes, which have previously been difficult to obtain via topology optimization.  The resulting thin-walled structures can provide valuable insights for designing thin-walled lightweight structures made of stamping, investment casting and composite manufacturing.
	\end{abstract}
\end{frontmatter}

\section{Introduction}
\label{sec:introduction}
\input{introduction.tex}

\section{Formulation}
\label{sec:formulation}
\input{formulation.tex}

\section{Adaptive meshing implementation}
\label{sec:adaptivemesh}
\input{adaptivemesh.tex}

\section{Numerical examples}
\label{sec:example}
\input{example.tex}

\section{Conclusion}
\label{sec:conclusion}
\input{conclusion.tex}

\input{reference.tex}
\end{document}

%% file: introduction.tex
Thin-walled structures are commonly seen in growing proportion of engineering applications ranging from automotive bodies, aircraft fuselages, and boat hulls. Cost and weight economy are the two primary factors that contribute to this growth. In addition, many manufacturing processes require (or prefer) the thin uniform thickness due to process limitations, e.g., stamping, investment casting, and composite manufacturing. The most basic requirement for thin-walled structures is the uniform thickness, which is small compared to its other dimensions.  While general thin-walled structures can contain branches and holes, a chosen manufacturing process may impose constraints on economically producing a component containing these features.

There are two major challenges to apply the density-based topology optimization method for designing 3D thin-walled structures. First, an effective geometric constraint is needed to guarantee the constant wall thickness. Secondly, for the thickness to be thin, the 3D design domain requires fine resolution discretization, which leads to high computational cost.

The control of the sizes of geometric features has been a long-studied topic in density-based topology optimization,  also known as the solid isotropic material with penalization (SIMP) method~\citep{bendsoe1989optimal,rozvany1992generalized,bendsoe2004topology}.  In particular, the minimum feature size control has been researched extensively, which is originally motivated by the concerns regarding mesh dependency and manufacturability. The spatial average-based low-pass filtering methods have been developed to regularize the sensitivity field~\citep{sigmund1997design} and the density field~\citep{bruns2001topology}. Such filters produce feature sizes under which density variation is not allowed. As a result, the minimum feature size can be indirectly controlled. \citet{poulsen2003new} proposed a computationally efficient integral constraint to impose the minimum feature size by checking the local density monotonicity. \citet{guest2004achieving} used nodal design variables and projection methods to achieve the minimum feature size while generating nearly black and white solutions. \citet{sigmund2007morphology} used morphology-based restriction scheme to generate nearly black and white designs while imposing both the minimum hole sizes and the minimum structural feature sizes. By applying geometric constraints to the filtered density field, \citet{zhou2015minimum} achieved the user-specified minimum feature size.

The maximum feature size control, on the other hand, has been recently investigated with motivations including channel size control in fluid filters, robustness against localized damage, fabrication concerns regarding thermal gradients and residual stresses.  \citet{guest2009imposing} used local constraints that imposed a minimum volume of void in each localized regions whose size would govern the maximum allowable feature size. \citet{zhang2014explicit} proposed to achieve the explicit and local control of both the minimum and maximum feature sizes by using the skeleton, which is a mathematical morphology concept that describes structural topologies. A band-pass filter projection method and a morphological-based method are proposed by~\citet{lazarov2017maximum} to impose the maximum feature size. In a projection-based framework, \citet{carstensen2018projection} presented a maximum feature size control scheme that could be applied to both material and void phases. \citet{fernandez2019aggregation} used an aggregation strategy that achieved the maximum feature size control using only a single constraint. While \citet{fernandez2019aggregation} did not attempt to achieve the uniform thickness, the maximum size constrained 3D structures were indeed close to thin-walled structures.

Most existing maximum feature size control methods require information about neighbor cells. Finding the neighbor cells requires a search for neighbors for each design point at every iteration. As discussed in ~\citet{lazarov2017maximum}, this is a very computationally expensive operation, especially, for 3D problems. Alternatively, the search operation can be performed as a preprocessing step, so that the neighbor cell information can be stored and reused. This approach, however, requires significantly more memory utilization. For adaptive mesh topology optimization, the preprocessing approach becomes impractical. It is noted topology optimization of infill lattice structures~\citep{wu2017minimum,wu2018infill,yi2019topology} shares a similar basic concept as the maximum feature size constrained topology optimization. The infill lattice patterns are often generated by enforcing local maximum allowable volume fraction constraints. When local volume fraction constraints are applied to localized regions in the size of the maximum allowable feature size, and the allowable local volume fraction is set to close to $1$, the local volume fraction constraint becomes equivalent to the maximum feature size constraint.

To design 3D thin-walled structures with a constant wall thickness manufactured by deep drawing, \citet{dienemann2017topology} used the mid surface and sensitivity penalization approach. Deep drawing is a sheet metal forming process that shapes sheet metal blanks by the punch mechanical action. While the optimized results are very suitable for the deep drawing application, this approach is not easily expandable to design general thin-walled structures because a punch direction is prescribed. \citet{clausen2015topology1,clausen2015topology2} used the spatial gradient of the density field in addition to a series of filtering and projection steps to achieve a uniform coating thickness. This approach is not deemed generalizable to 3D thin-walled structural design because infill materials are assumed and required to extract the coating shells. It is also not possible to have holes on the coating skin. \citet{zhang2016geometry,zhang2017stress,zhang2018geometry} used the geometric projection method to design structures made of uniform thickness plates. Each plate is explicitly represented by its shape and location design variables. While curved plates and holes can be attained by introducing additional design variables, the geometry attainable by this method is rather limited since they are constrained by the number and type of plates. 


To develop a general design method that does not assume any specific manufacturing processes, this paper adopts a uniform feature size control strategy, which enables a density-based topology optimization for designing general 3D thin-walled structures containing branches and holes. The uniform thin thickness is achieved by simultaneously constraining the minimum and maximum feature sizes while narrowing the gap between them. It is noted that a similar concept has been applied to the moving morphable components topology optimization method~\citep{niu2018equal}, to generate equal-width truss-like 2D designs. The proposed method for achieving uniform feature sizes is based on the Helmholtz PDE filtering approach~\citep{lazarov2011filters,kawamoto2011heaviside}, which has originally applied for the minimum feature size control. This paper proposes a method that also controls the maximum feature size by using an additional Helmholtz PDE filter and an aggregated constraint. The PDE-based method was chosen since they do not require information about neighbor cells, which is especially convenient for problems with fine unstructured mesh and adaptive meshing.

To realize high-resolution 3D topology optimization, different strategies have been investigated, based on, for example, hardware acceleration~\citep{wu2016system}, parallel computation~\citep{aage2015topology}, and the efficient use of effective elements~\citep{liu2018narrow} . Though these latest developments have successfully pushed the density-based topology optimization to handle billions of 3D voxels, they often require massive hardware resources and/or extensive low-level programming, which can be overwhelming for topology optimization researchers in the early explorative development stage. Following the concept of efficient use of effective elements, this paper presents an adaptive meshing strategy that can be implemented in high-level programming platforms, {\em e.g.}, COMSOL Multiphysics. The proposed adaptive meshing strategy saves significant computation that otherwise is wasted in large void regions that appear in problems with low material-void ratios like 3D thin-walled structures.  Several numerical examples are presented to demonstrate the effectiveness of the proposed method in designing general 3D thin-walled structures with uniform wall thickness, branches, and holes. It is also demonstrated that an adaptive meshing scheme realizes the thin thickness and fine geometric features in the final design using only standard desktop PCs.

The rest of the paper is organized as follows. Section~\ref{sec:formulation} presents the PDE-based minimum and maximum feature size control methods and the thin-walled structural topology optimization formulation. Section~\ref{sec:adaptivemesh} details the proposed adaptive mesh scheme and its implementation. Section~\ref{sec:example} discusses several numerical examples. Finally, Section~\ref{sec:conclusion} summarizes the current study and opportunities for future research.

%% file: formulation.tex
The design field regularization and the PDE-based minimum feature size control follow the formulation discussed in~\citet{kawamoto2011heaviside}. In a prescribed, fixed design domain $D$, a characteristic function $\chi$ is defined to describe the material domain $\Omega_d$ to be optimized:
\begin{equation}
	\chi(\mathbf{x}) =
	\begin{cases}
	~0 &\mbox{ for ~ } 
	\forall\mathbf{x}\in D\setminus\Omega_d \\
	~1 &\mbox{ for ~ }  \forall\mathbf{x}\in \Omega_d\\
	\end{cases},
\end{equation}
where $\mathbf{x}$ stands for a design point in $D$. The characteristic function $\chi(\mathbf{x})$ is defined by a scalar function $\phi$ and a Heaviside function $H$ such that:
\begin{equation}
	\chi(\mathbf{x}) =H\left(\phi(\mathbf{x})\right) =
	\begin{cases}
	~0 &\mbox{ for ~}  \forall\mathbf{x}\in D\setminus \Omega_d\\
	~1 &\mbox{ for ~}  \forall\mathbf{x}\in\Omega_d
	\end{cases}.
\end{equation}
Based on the definition, the scalar function $\phi$ as a design field can take any real values. For a practical optimization formulation, $\phi$ can be bounded between $-1$ and $1$.

As in the original Helmholtz PDE method \citep{lazarov2011filters,kawamoto2011heaviside}, the first Helmholtz PDE is introduced to regularize $\phi$:
\begin{equation}
	-\ubar{r}^2\nabla^2\tilde{\phi}+\tilde{\phi}=\phi,
\label{eq:pde1}
\end{equation}
where $\tilde{\phi}$ is the regularized field after filtering. Parameter $\ubar{r}$ is the minimum feature radius, which governs the minimum geometric feature size. According to~\citet{clausen2015topology1}, the relation between $\ubar{r}$ and the filter radius $\ubar{R}$, often seen in standard spatial average-based filtering methods, which governs the minimum geometric feature size is:
\begin{equation}
	\ubar{r} = \dfrac{\ubar{R}}{2\sqrt{3}}.
	\label{eq:relation}
\end{equation}
The density field $\rho$ can then be defined by an additional smoothed Heaviside function $\tilde{H}$:
\begin{equation}
	\rho = \tilde{H}(\tilde{\phi}).
	\label{eq:heaviside1}
\end{equation}

After the series of regularization from $\phi$ to $\rho$, the resulting density field $\rho$ is bounded between 0 and 1, which describes the design of topology and will be used for structural performance evaluation. The resulting topology design should have no geometric features smaller than the minimum feature size $\ubar{R}$.

\subsection{PDE-based maximum feature size constraint}
By applying the third smoothed Heaviside function $\bar{H}$ (different from the aforementioned $H$ and $\tilde{H}$) to the regularized field $\tilde{\phi}$, the resulting field $\tilde{\rho}$ (different from the aforementioned $\rho$) can be bounded between 0 and 1 as follows:
\begin{equation}
	\tilde{\rho} = \bar{H}(\tilde{\phi}).
	\label{eq:heaviside2}
\end{equation}
The independent definitions of $\rho$ and $\tilde{\rho}$ separate the physics analysis domain and the geometric evaluation domain. During the course of optimization, a continuation scheme can be applied to Equation~(\ref{eq:heaviside1}) to avoid early local solution trap while a consistently narrowing Heaviside bandwidth can be applied to Equation~(\ref{eq:heaviside2}) to ensure the satisfaction of geometric constraints at all iterations.

An additional Helmholtz PDE with a different filter radius $\bar{r}$ setting (larger than the aforementioned $\ubar{r}$) is applied to $\tilde{\rho}$ as follows,
\begin{equation}
	-\bar{r}^2\nabla^2\bar{\rho}+\bar{\rho}=\tilde{\rho}.
	\label{eq:pde2}
\end{equation}
Similar to Equation~(\ref{eq:relation}), $\bar{r}$ can be linked with the actual geometric feature size $\bar{R}$. Theoretically, for a topology without any geometric features larger than $\bar{R}$, the diffused term $\bar{\rho}$ will always be strictly smaller than $1$. Otherwise, $\bar{\rho} \geq 1$ will occur at regions with geometric features larger than $\bar{R}$. Therefore, by constraining the maximum allowable value on $\bar{\rho}$, one can indirectly control the maximum allowable feature size $\bar{R}$ in the optimized topology.

Conceptually, the proposed PDE-based method shares a similar trait to previously reported work on the maximum feature size control method~\citep{guest2009imposing} and the lattice infill method~\citep{wu2017minimum,wu2018infill}. The primary advantage of the proposed PDE-based method is that the information about neighbor cells is not required, which makes it especially appealing to problems with fine unstructured mesh and adaptive meshing.

To effectively constrain the maximum value of $\bar{\rho}$, \citet{guest2009imposing} used many local constraints and \citet{wu2017minimum,wu2018infill} used a single P-norm aggregated constraint. This paper follows a Heaviside projection based aggregation method, which has been previously applied to the stress-constrained topology optimization~\citep{wang2018heaviside}. The integral constraint is formulated as follows:
\begin{equation}
	\int_{D}{\hat{H}\left(\bar{\rho} - \beta, h\right) \left(\frac{\bar{\rho}}{\beta}\right)^{\eta} }\dd{\Omega}-\epsilon^{\ast} \leqslant 0,
	\label{eq:constraint}
\end{equation}
where $\beta$ is the prescribed maximum allowable $\bar{\rho}$, $h$ is the smoothed Heaviside bandwidth parameter, $\eta$ is the penalty factor, $\epsilon^{\ast}$ is the integral bound that determines the allowable violation of the maximum feature size constraint, and $\hat{H}$ is a smoothed Heaviside function with a continuous second derivative:
\begin{equation}
	\hat{H}\left( x, h \right) =
	\begin{cases}
	~0 &\mbox{ for ~ } 	x < -h \\
	~\frac{1}{2} + \frac{15}{16} \left(x/h\right) - \frac{5}{8}\left(x/h\right)^3 + \frac{3}{16}\left(x/h\right)^5 &\mbox{ for ~ }  -h \leq	x \leq h \\
	~1 &\mbox{ for ~ }  x > h \\
	\end{cases}.
\end{equation}

In order to precisely bound the maximum feature size to be $\bar{R}$, $\beta$ should be set to a value smaller than yet very close to $1$, and $h$ should be set to an infinitesimal value. However, this setting is not practical for the sensitivity-driven numerical optimization due to the lack of smoothness in the Heaviside function $\hat{H}$. To resolve this numerical challenge, a filter radius larger than $\bar{r}$ should be used in Equation~(\ref{eq:pde2}) so that moderate values for $\beta$ and $h$ can be used.

\subsection{Uniform feature size control}
\label{subsec:uniform}
With the appropriate settings of $\ubar{r}$ and $\bar{r}$ (or $\ubar{R}$ and $\bar{R}$), both the minimum and maximum feature sizes can be explicitly controlled. In order to achieve the uniform feature size control, both feature size radius parameters can be set as identical. However, this will lead to the numerical challenge for the maximum feature size constraint as discussed in the previous section. To investigate the appropriate settings for feature size radius parameters $\ubar{r}$ and $\bar{r}$ and smoothed Heaviside function parameters $\beta$ and $h$, several numerical tests are conducted for the demonstration purpose.

\begin{figure}
    \centering
    \scriptsize
	\def\svgwidth{0.9\linewidth}
	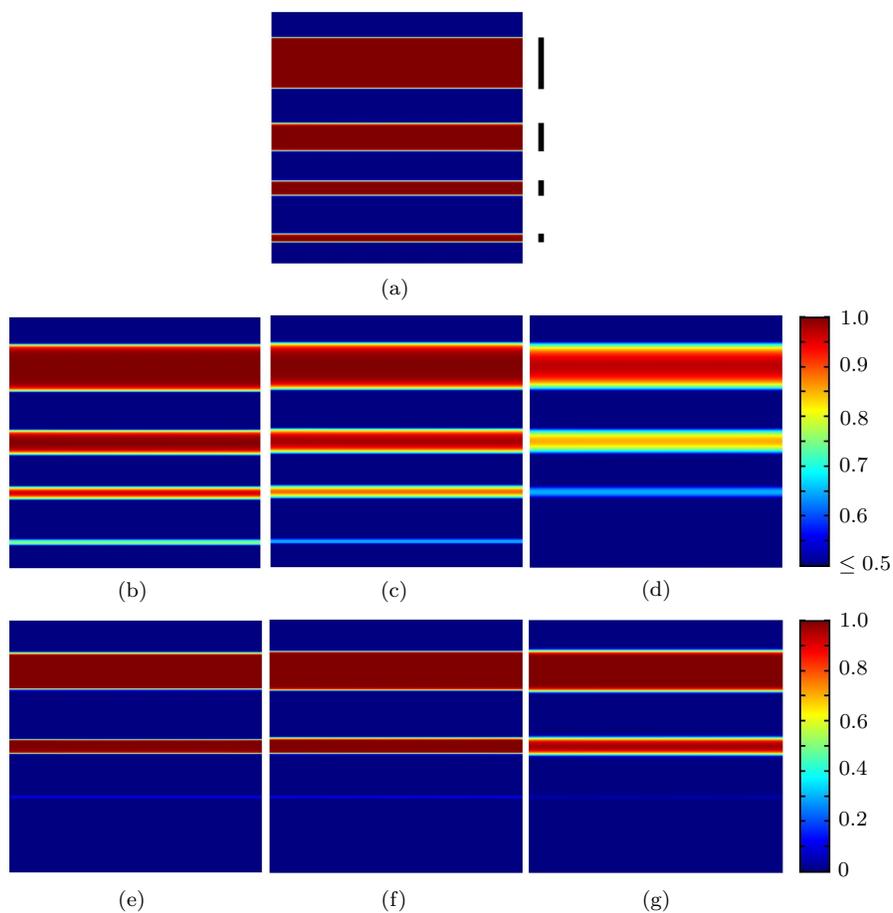
	\caption{Demonstration of the relationship between feature size radii and the parameters for the  smoothed Heaviside function. (a) Four strips (representation of $\tilde{\rho}$) with different feature sizes: $4\ubar{R}$, $2\ubar{R}$, $\ubar{R}$ and $0.5\ubar{R}$ (from top to bottom). (b)-(d) The diffused fields (representation of $\bar{\rho}$) with different filter radius settings: $\bar{R} = 1.5\ubar{R}$, $\bar{R} = 2\ubar{R}$, $\bar{R} = 4\ubar{R}$. (e)-(g) Maximum feature size violation detection (representation of $\hat{H}\left(\bar{\rho} - \beta, h\right)$) with different smoothed Heaviside function settings $(\beta, h)$: $(0.97, 0.015)$, $(0.90, 0.05)$, $(0.75, 0.2)$.}
    \label{fig:demo}
\end{figure}

As seen in Figure~\ref{fig:demo}(a), four strips with different feature sizes are artificially created as the representation of $\tilde{\rho}$. Since a tight bandwidth parameter is recommended in Equation~(\ref{eq:heaviside2}), $\tilde{\rho}$ will always be almost 1 or 0. The third row (from top to bottom) indicates a strip governed by the assumed minimum feature size $\ubar{R}$. The top first two rows are strips with larger feature sizes $4\ubar{R}$ and $2\ubar{R}$. The bottom row indicates a strip with a feature size smaller than $\ubar{R}$, which theoretically should have been avoided after the first Helmholtz PDE and Heaviside projection. By specifying different filter radius $\bar{r}$ parameters in Equation~(\ref{eq:pde2}), Figures~\ref{fig:demo}(b)-(d) show three diffused fields with equivalent geometric feature radius $\bar{R}$ settings of $1.5\ubar{R}$, $2\ubar{R}$ and $4\ubar{R}$ respectively. It is noted with the increase of $\bar{r}$, the maximum values at the center of strips decrease. The strategy to avoid the formation of any geometric features larger than the prescribed maximum feature size $\bar{R}$ is to identify and penalize them in the integral constraint of Equation~(\ref{eq:constraint}). Figures~\ref{fig:demo}(e)-(g) demonstrate that with the appropriate settings for the smoothed Heaviside function $\hat{H}$ in Equation~(\ref{eq:constraint}), geometric features smaller than $\bar{R}$ can be filtered out while geometric features larger than $\bar{R}$ can be identified. With the penalization of identified large geometric features and an infinitesimal allowable constraint violation setting $\epsilon^{\ast}$, the resulting topology is expected to have no geometric features larger than the prescribed $\bar{R}$.

\begin{figure}
    \centering
    \scriptsize
	\def\svgwidth{0.9\linewidth}
	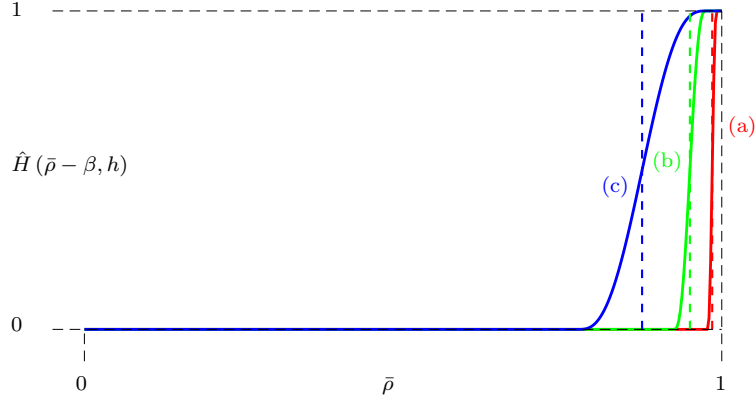
	\caption{Smoothed Heaviside functions with different parameters $(\beta, h)$: (a) $(0.97, 0.015)$ (b) $(0.90, 0.05)$ (c) $(0.75, 0.2)$.}
\label{fig:heaviside}
\end{figure}

The three smoothed Heaviside function profiles with different settings to generate Figures~\ref{fig:demo}(e)-(g) are plotted in Figure~\ref{fig:heaviside}. It can be seen that the function smoothness improves with smaller $\beta$ and larger $h$ settings. Such function smoothness improvement is also only available with wider gap settings between $\ubar{r}$ and $\bar{r}$ (or $\ubar{R}$ and $\bar{R}$). It is noted while the numerical tests can provide some rule-of-thumb knowledge regarding the relationship between filter radius parameters and smoothed Heaviside function settings in order to achieve the uniform feature size control, the actual settings may vary in different topology optimization problems.

\subsection{Optimization problem}
The overall topology optimization problem with a volume fraction constraint and the proposed uniform feature size constraint for designing thin-walled structures is summarized as follows:
\begin{equation}
	\begin{aligned}
	& \underset{\substack{\phi}}{\text{minimize}}
	&& F \\
	&  \text{\text{subject to}}
	&& g_1:=\int_{D}{\frac{\rho}{V_0}}\dd{\Omega}-V^{\ast} \leqslant 0 \\
	&&& g_2:=\int_{D}{\hat{H}\left(\bar{\rho} - \beta, h\right) \left(\frac{\bar{\rho}}{\beta}\right)^{\eta} }\dd{\Omega}-\epsilon^{\ast} \leqslant 0\\
	&&& \phi\in[-1,1]^D \\
	&&& \text{Equilibrium equations} \\
	&&& \text{Material interpolation}
	\end{aligned},
\label{eq:optimization}
\end{equation}
where $\phi$ is the design field ranging between $-1$ and $1$. The transformations from $\phi$ to $\rho$ and $\bar{\rho}$ are detailed in Equations~(\ref{eq:pde1})-(\ref{eq:pde2}). Function $F$ is the structural performance objective. Function $g_1$ is the volume fraction constraint where $V_0$ is the total volume of the design domain, and $V^{\ast}$ is the prescribed maximum allowable volume fraction. Function $g_2$ is the uniform feature size constraint detailed in Section~\ref{subsec:uniform}.

For structural compliance minimization problems, $F$ is:
\begin{equation}
	F(\mathbf{u})=\int_D\frac{1}{2}\boldsymbol{\sigma}^{\intercal} \boldsymbol{\epsilon}\dd\Omega,
\end{equation}
and the static equilibrium equations are:
\begin{equation}
	\begin{array}{rcll}
	\nabla\cdot\boldsymbol{\sigma}&=&\mathbf{0}&\mbox{ in } D\\
	\mathbf{u}&=&\mathbf{0} &\mbox{ on } \Gamma_d\\
	\boldsymbol{\sigma}\cdot\mathbf{n}&=&\mathbf{t}&\mbox{ on } \Gamma_n\\
	\end{array},
\end{equation}
where $\boldsymbol{\sigma}=\mathbf{C}\cdot\boldsymbol{\epsilon}(\mathbf{u})$ is the stress field. $\boldsymbol{\epsilon}(\mathbf{u})$ is the strain field. $\mathbf{C}$ is the elasticity tensor. $\Gamma_d$ is the Dirichlet boundary. $\Gamma_n$ is the Neumann boundary.

The elastic tensor is obtained by the SIMP material interpolation scheme as follows:
\begin{equation}
	\mathbf{C} = \rho^{P}\mathbf{C}_0,
\end{equation}
where $\mathbf{C}_0$ is the full elasticity tensor. $\rho$ is the regularized material density. $P$ is the penalization parameter for the SIMP power law.

%% file: figures/demo.pdf_tex
\begingroup%
  \makeatletter%
  \providecommand\color[2][]{%
    \errmessage{(Inkscape) Color is used for the text in Inkscape, but the package 'color.sty' is not loaded}%
    \renewcommand\color[2][]{}%
  }%
  \providecommand\transparent[1]{%
    \errmessage{(Inkscape) Transparency is used (non-zero) for the text in Inkscape, but the package 'transparent.sty' is not loaded}%
    \renewcommand\transparent[1]{}%
  }%
  \providecommand\rotatebox[2]{#2}%
  \newcommand*\fsize{\dimexpr\f@size pt\relax}%
  \newcommand*\lineheight[1]{\fontsize{\fsize}{#1\fsize}\selectfont}%
  \ifx\svgwidth\undefined%
    \setlength{\unitlength}{468bp}%
    \ifx\svgscale\undefined%
      \relax%
    \else%
      \setlength{\unitlength}{\unitlength * \real{\svgscale}}%
    \fi%
  \else%
    \setlength{\unitlength}{\svgwidth}%
  \fi%
  \global\let\svgwidth\undefined%
  \global\let\svgscale\undefined%
  \makeatother%
  \begin{picture}(1,1)%
    \lineheight{1}%
    \setlength\tabcolsep{0pt}%
    \put(0,0){\includegraphics[width=\unitlength,page=1]{demo.pdf}}%
    \put(0.90363534,0.65030489){\color[rgb]{0,0,0}\makebox(0,0)[lt]{\lineheight{1.25}\smash{\begin{tabular}[t]{l}$1.0$\end{tabular}}}}%
    \put(0.90295891,0.59908277){\color[rgb]{0,0,0}\makebox(0,0)[lt]{\lineheight{1.25}\smash{\begin{tabular}[t]{l}$0.9$\end{tabular}}}}%
    \put(0.90339467,0.54191027){\color[rgb]{0,0,0}\makebox(0,0)[lt]{\lineheight{1.25}\smash{\begin{tabular}[t]{l}$0.8$\end{tabular}}}}%
    \put(0.90271824,0.49068806){\color[rgb]{0,0,0}\makebox(0,0)[lt]{\lineheight{1.25}\smash{\begin{tabular}[t]{l}$0.7$\end{tabular}}}}%
    \put(0.90258525,0.43613211){\color[rgb]{0,0,0}\makebox(0,0)[lt]{\lineheight{1.25}\smash{\begin{tabular}[t]{l}$0.6$\end{tabular}}}}%
    \put(0.90190881,0.38491008){\color[rgb]{0,0,0}\makebox(0,0)[lt]{\lineheight{1.25}\smash{\begin{tabular}[t]{l}$\leq0.5$\end{tabular}}}}%
    \put(0,0){\includegraphics[width=\unitlength,page=2]{demo.pdf}}%
    \put(0.40827448,0.6825453){\color[rgb]{0,0,0}\makebox(0,0)[lt]{\lineheight{1.25}\smash{\begin{tabular}[t]{l}(a)\end{tabular}}}}%
    \put(0.1246934,0.35619925){\color[rgb]{0,0,0}\makebox(0,0)[lt]{\lineheight{1.25}\smash{\begin{tabular}[t]{l}(b)\end{tabular}}}}%
    \put(0.4081572,0.35619925){\color[rgb]{0,0,0}\makebox(0,0)[lt]{\lineheight{1.25}\smash{\begin{tabular}[t]{l}(c)\end{tabular}}}}%
    \put(0.68928973,0.35700867){\color[rgb]{0,0,0}\makebox(0,0)[lt]{\lineheight{1.25}\smash{\begin{tabular}[t]{l}(d)\end{tabular}}}}%
    \put(-0.11904762,0.86996343){\color[rgb]{0,0,0}\makebox(0,0)[lt]{\begin{minipage}{1.29349828\unitlength}\raggedright \end{minipage}}}%
    \put(0,0){\includegraphics[width=\unitlength,page=3]{demo.pdf}}%
    \put(0.90311843,0.32346485){\color[rgb]{0,0,0}\makebox(0,0)[lt]{\lineheight{1.25}\smash{\begin{tabular}[t]{l}$1.0$\end{tabular}}}}%
    \put(0.90244199,0.27224272){\color[rgb]{0,0,0}\makebox(0,0)[lt]{\lineheight{1.25}\smash{\begin{tabular}[t]{l}$0.8$\end{tabular}}}}%
    \put(0.90287776,0.21506921){\color[rgb]{0,0,0}\makebox(0,0)[lt]{\lineheight{1.25}\smash{\begin{tabular}[t]{l}$0.6$\end{tabular}}}}%
    \put(0.90220133,0.16384552){\color[rgb]{0,0,0}\makebox(0,0)[lt]{\lineheight{1.25}\smash{\begin{tabular}[t]{l}$0.4$\end{tabular}}}}%
    \put(0.90206833,0.109288){\color[rgb]{0,0,0}\makebox(0,0)[lt]{\lineheight{1.25}\smash{\begin{tabular}[t]{l}$0.2$\end{tabular}}}}%
    \put(0.9013919,0.05485931){\color[rgb]{0,0,0}\makebox(0,0)[lt]{\lineheight{1.25}\smash{\begin{tabular}[t]{l}$0$\end{tabular}}}}%
    \put(0.125359,0.02215263){\color[rgb]{0,0,0}\makebox(0,0)[lt]{\lineheight{1.25}\smash{\begin{tabular}[t]{l}(e)\end{tabular}}}}%
    \put(0.40882282,0.02215263){\color[rgb]{0,0,0}\makebox(0,0)[lt]{\lineheight{1.25}\smash{\begin{tabular}[t]{l}(f)\end{tabular}}}}%
    \put(0.6899553,0.02296206){\color[rgb]{0,0,0}\makebox(0,0)[lt]{\lineheight{1.25}\smash{\begin{tabular}[t]{l}(g)\end{tabular}}}}%
  \end{picture}%
\endgroup%

%% file: figures/heaviside.pdf_tex
\begingroup%
  \makeatletter%
  \providecommand\color[2][]{%
    \errmessage{(Inkscape) Color is used for the text in Inkscape, but the package 'color.sty' is not loaded}%
    \renewcommand\color[2][]{}%
  }%
  \providecommand\transparent[1]{%
    \errmessage{(Inkscape) Transparency is used (non-zero) for the text in Inkscape, but the package 'transparent.sty' is not loaded}%
    \renewcommand\transparent[1]{}%
  }%
  \providecommand\rotatebox[2]{#2}%
  \newcommand*\fsize{\dimexpr\f@size pt\relax}%
  \newcommand*\lineheight[1]{\fontsize{\fsize}{#1\fsize}\selectfont}%
  \ifx\svgwidth\undefined%
    \setlength{\unitlength}{468bp}%
    \ifx\svgscale\undefined%
      \relax%
    \else%
      \setlength{\unitlength}{\unitlength * \real{\svgscale}}%
    \fi%
  \else%
    \setlength{\unitlength}{\svgwidth}%
  \fi%
  \global\let\svgwidth\undefined%
  \global\let\svgscale\undefined%
  \makeatother%
  \begin{picture}(1,0.53846154)%
    \lineheight{1}%
    \setlength\tabcolsep{0pt}%
    \put(-0.11904762,0.86996345){\color[rgb]{0,0,0}\makebox(0,0)[lt]{\begin{minipage}{1.29349826\unitlength}\raggedright \end{minipage}}}%
    \put(0,0){\includegraphics[width=\unitlength,page=1]{heaviside.pdf}}%
    \put(0.16483518,0.04739002){\color[rgb]{0,0,0}\makebox(0,0)[lt]{\lineheight{1.25}\smash{\begin{tabular}[t]{l}$0$\end{tabular}}}}%
    \put(0.85397125,0.04714344){\color[rgb]{0,0,0}\makebox(0,0)[lt]{\lineheight{1.25}\smash{\begin{tabular}[t]{l}$1$\end{tabular}}}}%
    \put(0.09535197,0.11054293){\color[rgb]{0,0,0}\makebox(0,0)[lt]{\lineheight{1.25}\smash{\begin{tabular}[t]{l}$0$\end{tabular}}}}%
    \put(0.09539679,0.45031764){\color[rgb]{0,0,0}\makebox(0,0)[lt]{\lineheight{1.25}\smash{\begin{tabular}[t]{l}$1$\end{tabular}}}}%
    \put(0.73154598,0.26110559){\color[rgb]{0,0,1}\makebox(0,0)[lt]{\lineheight{1.25}\smash{\begin{tabular}[t]{l}(c)\end{tabular}}}}%
    \put(0.78608017,0.2874253){\color[rgb]{0,1,0}\makebox(0,0)[lt]{\lineheight{1.25}\smash{\begin{tabular}[t]{l}(b)\end{tabular}}}}%
    \put(0.8677987,0.32541118){\color[rgb]{1,0,0}\makebox(0,0)[lt]{\lineheight{1.25}\smash{\begin{tabular}[t]{l}(a)\end{tabular}}}}%
    \put(0.49510378,0.04734197){\color[rgb]{0,0,0}\makebox(0,0)[lt]{\lineheight{1.25}\smash{\begin{tabular}[t]{l}$\bar{\rho}$\end{tabular}}}}%
    \put(0.09566274,0.28352372){\color[rgb]{0,0,0}\makebox(0,0)[lt]{\lineheight{1.25}\smash{\begin{tabular}[t]{l}$\hat{H}\left(\bar{\rho} - \beta, h\right)$\end{tabular}}}}%
  \end{picture}%
\endgroup%

%% file: adaptivemesh.tex
In the scope of saving computational effort otherwise wasted in void regions (i.e., low density $\rho$) during topology optimization, past research has two primary adaptive strategies. The first class adaptively removes and reintroduces elements~\citep{bruns2003element, guest2010reducing, liu2018narrow}. As a result, the design domain changes during the course of optimization. In addition to the computational saving due to the removal of degrees of freedom, this adaptive strategy has also demonstrated superior performance in fluid problems, which can avoid flow seepage through the solid material phase~\citep{behrou2019adaptive}. Some drawbacks of removing elements from the design domain are, for example, the full-space expensive computation at the beginning of the optimization, and the robustness of reintroducing elements in void regions. The second class refines the mesh based on a prescribed local error indicator~\citep{maute1995adaptive,bruggi2011fully,wang2014adaptive,yamasaki2015consistent,lambe2018topology,nguyen2017polytree,chin2018efficient,de2018adaptive,wu2018continuous,baiges2019large}. While the mesh changes during the course of optimization, the design domain is fixed. In general, the mesh is refined in regions of interest defined by the local error indicator while coarsened in non-critical regions. The primary advantage of mesh refinement methods over the element removal methods is its ability to allow structural member appearance anywhere in the design domain in all optimization iterations. The adaptive meshing scheme implemented in this paper follows the mesh refinement strategy.

The proposed mesh refinement criterion is based on a prescribed error indicator $w$, which is defined with respect to the density field $\rho$ at the mesh refinement iteration:
\begin{equation}
    w = \rho \left( 1-\rho+\alpha \right),
\end{equation}
where $\alpha$ is a parameter ranging between $0$ and $1$, governing the coarsening effect in established geometries with high density values.

\begin{figure}
    \centering
    \scriptsize
	\def\svgwidth{0.9\linewidth}
	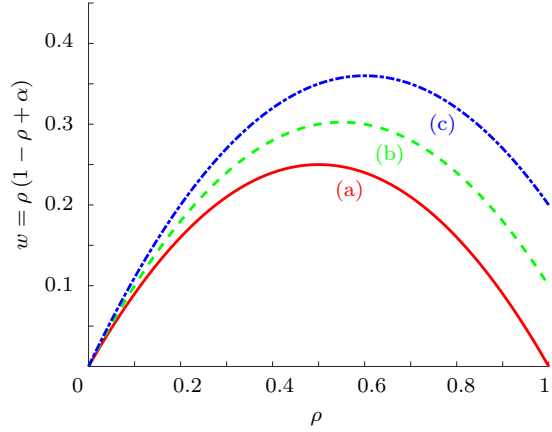
	\caption{The adaptive mesh error indicator profiles with different $\alpha$ settings. (a) $\alpha = 0$ (b) $\alpha = 0.1$ (c) $\alpha = 0.2$}
\label{fig:weight}
\end{figure}

The relationship between the density $\rho$ and the error indicator $w$ is plotted in Figure~\ref{fig:weight} with three different $\alpha$ settings. In regions of the design domain, where there is no material, i.e. $\rho \approx 0$, a coarse mesh is used. In regions of the design domain, where the structural layout is blurry, i.e. $ 0 < \rho < 1$, a fine mesh is generated. Since new holes can emerge in regions with clear structural members, i.e. $\rho \approx 1$, certain refinement level should be kept. $\alpha = 1$ indicates that the finest mesh happens in $\rho \approx 1$ regions. To assign more elements to exploitative regions during the course of optimization, a smaller $\alpha$ value can be set. This will also lead to structural boundary refinement at the end of optimization. It is noted that the error indicator profile can be customized based on different problems. A continuation scheme can also be applied to further fine tune the adaptive mesh refinement behavior during the course of optimization.

Based on COMSOL Multiphysics, this paper utilizes a versatile high-level programming-language implementation for adaptive mesh topology optimization. COMSOL 5.3a or later is required for the implementation of adaptive mesh based on location dependent error indicators (i.e., a field variable). COMSOL LiveLink for MATLAB is used for the integration of adaptive mesh into topology optimization iterations. The mesh initialization refinement method based on free tetrahedral elements is implemented in this paper. The element growth rate is kept constant throughout the course of optimization. By setting an element growth rate larger than $1$ (or smaller than $1$), the number of elements will increase (or decrease) at each mesh refinement iteration. For more details about the adaptive mesh setting customization, readers are referred to the COMSOL documentation.

%% file: figures/weight.pdf_tex
\begingroup%
  \makeatletter%
  \providecommand\color[2][]{%
    \errmessage{(Inkscape) Color is used for the text in Inkscape, but the package 'color.sty' is not loaded}%
    \renewcommand\color[2][]{}%
  }%
  \providecommand\transparent[1]{%
    \errmessage{(Inkscape) Transparency is used (non-zero) for the text in Inkscape, but the package 'transparent.sty' is not loaded}%
    \renewcommand\transparent[1]{}%
  }%
  \providecommand\rotatebox[2]{#2}%
  \newcommand*\fsize{\dimexpr\f@size pt\relax}%
  \newcommand*\lineheight[1]{\fontsize{\fsize}{#1\fsize}\selectfont}%
  \ifx\svgwidth\undefined%
    \setlength{\unitlength}{468bp}%
    \ifx\svgscale\undefined%
      \relax%
    \else%
      \setlength{\unitlength}{\unitlength * \real{\svgscale}}%
    \fi%
  \else%
    \setlength{\unitlength}{\svgwidth}%
  \fi%
  \global\let\svgwidth\undefined%
  \global\let\svgscale\undefined%
  \makeatother%
  \begin{picture}(1,0.52307694)%
    \lineheight{1}%
    \setlength\tabcolsep{0pt}%
    \put(-0.11904762,0.86996347){\color[rgb]{0,0,0}\makebox(0,0)[lt]{\begin{minipage}{1.29349829\unitlength}\raggedright \end{minipage}}}%
    \put(0,0){\includegraphics[width=\unitlength,page=1]{weight.pdf}}%
    \put(0.26128476,0.04238023){\color[rgb]{0,0,0}\makebox(0,0)[lt]{\lineheight{1.25}\smash{\begin{tabular}[t]{l}$0$\end{tabular}}}}%
    \put(0.37029876,0.04235195){\color[rgb]{0,0,0}\makebox(0,0)[lt]{\lineheight{1.25}\smash{\begin{tabular}[t]{l}$0.2$\end{tabular}}}}%
    \put(0.46959204,0.04227672){\color[rgb]{0,0,0}\makebox(0,0)[lt]{\lineheight{1.25}\smash{\begin{tabular}[t]{l}$0.4$\end{tabular}}}}%
    \put(0.56889307,0.04219058){\color[rgb]{0,0,0}\makebox(0,0)[lt]{\lineheight{1.25}\smash{\begin{tabular}[t]{l}$0.6$\end{tabular}}}}%
    \put(0.66499734,0.04237764){\color[rgb]{0,0,0}\makebox(0,0)[lt]{\lineheight{1.25}\smash{\begin{tabular}[t]{l}$0.8$\end{tabular}}}}%
    \put(0.76437823,0.04240518){\color[rgb]{0,0,0}\makebox(0,0)[lt]{\lineheight{1.25}\smash{\begin{tabular}[t]{l}$1$\end{tabular}}}}%
    \put(0.23181062,0.15202134){\color[rgb]{0,0,0}\makebox(0,0)[lt]{\lineheight{1.25}\smash{\begin{tabular}[t]{l}$0.1$\end{tabular}}}}%
    \put(0.23171452,0.23858572){\color[rgb]{0,0,0}\makebox(0,0)[lt]{\lineheight{1.25}\smash{\begin{tabular}[t]{l}$0.2$\end{tabular}}}}%
    \put(0.23181062,0.32516655){\color[rgb]{0,0,0}\makebox(0,0)[lt]{\lineheight{1.25}\smash{\begin{tabular}[t]{l}$0.3$\end{tabular}}}}%
    \put(0.23179135,0.41463731){\color[rgb]{0,0,0}\makebox(0,0)[lt]{\lineheight{1.25}\smash{\begin{tabular}[t]{l}$0.4$\end{tabular}}}}%
    \put(0.5446126,0.25412612){\color[rgb]{1,0,0}\makebox(0,0)[lt]{\lineheight{1.25}\smash{\begin{tabular}[t]{l}(a)\end{tabular}}}}%
    \put(0.58732885,0.29322152){\color[rgb]{0,1,0}\makebox(0,0)[lt]{\lineheight{1.25}\smash{\begin{tabular}[t]{l}(b)\end{tabular}}}}%
    \put(0.64562459,0.32523957){\color[rgb]{0,0,1}\makebox(0,0)[lt]{\lineheight{1.25}\smash{\begin{tabular}[t]{l}(c)\end{tabular}}}}%
    \put(0.51806887,0.01177115){\color[rgb]{0,0,0}\makebox(0,0)[lt]{\lineheight{1.25}\smash{\begin{tabular}[t]{l}$\rho$\end{tabular}}}}%
    \put(0.21267673,0.19557867){\color[rgb]{0,0,0}\rotatebox{90}{\makebox(0,0)[lt]{\lineheight{1.25}\smash{\begin{tabular}[t]{l}$w=\rho\left( 1-\rho+\alpha\right)$\end{tabular}}}}}%
  \end{picture}%
\endgroup%

%% file: example.tex
\begin{figure}
    \centering
    \scriptsize
	\def\svgwidth{0.9\linewidth}
	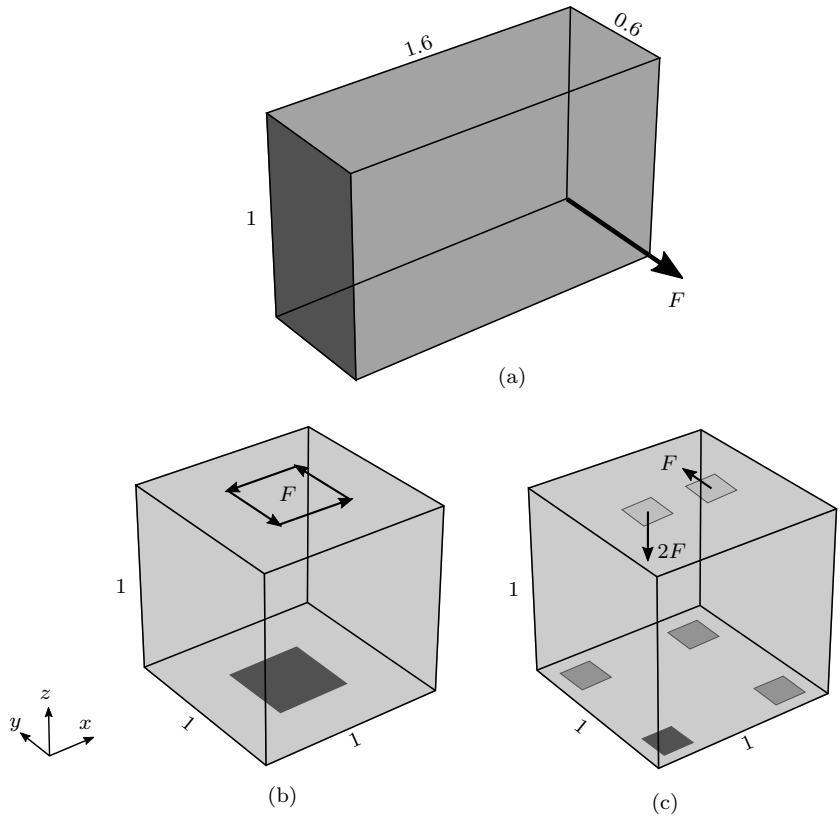
	\caption{Design domain and boundary condition settings for the examples of (a) sheared beam, (b) twisted ball, and (c) multi-direction loaded cube.}
\label{fig:3d_bc}
\end{figure}

Several 3D examples are presented to demonstrate the proposed thin-walled structural topology optimization method. Their design domain and boundary condition settings are summarized in Figure~\ref{fig:3d_bc}. The first two examples used a standard desktop PC (CPU: Xeon E3-1241 v3 3.5GHz; RAM: 16 GB). The third higher resolution example used a higher-end desktop PC (CPU: Xeon E5-1680 v4 3.4 GHz; RAM: 128 GB). COMSOL Multiphysics is used to solve PDEs, perform sensitivity analysis, and implement adaptive meshing. COMSOL LiveLink for MATLAB is used to integrate COMSOL solvers into the MATLAB controlled optimization loop. The nonlinear constrained optimization problem is solved by the Method of Moving Asymptotes~\citep{svanberg1987method}. The maximum allowable number of optimization iterations is $100$. The optimization terminates either the change of the objective function and design variables are below prescribed values or the maximum allowable number of iterations is reached.

\subsection{Sheared beam}
\begin{figure}
    \centering
    \scriptsize
	\def\svgwidth{0.85\linewidth}
	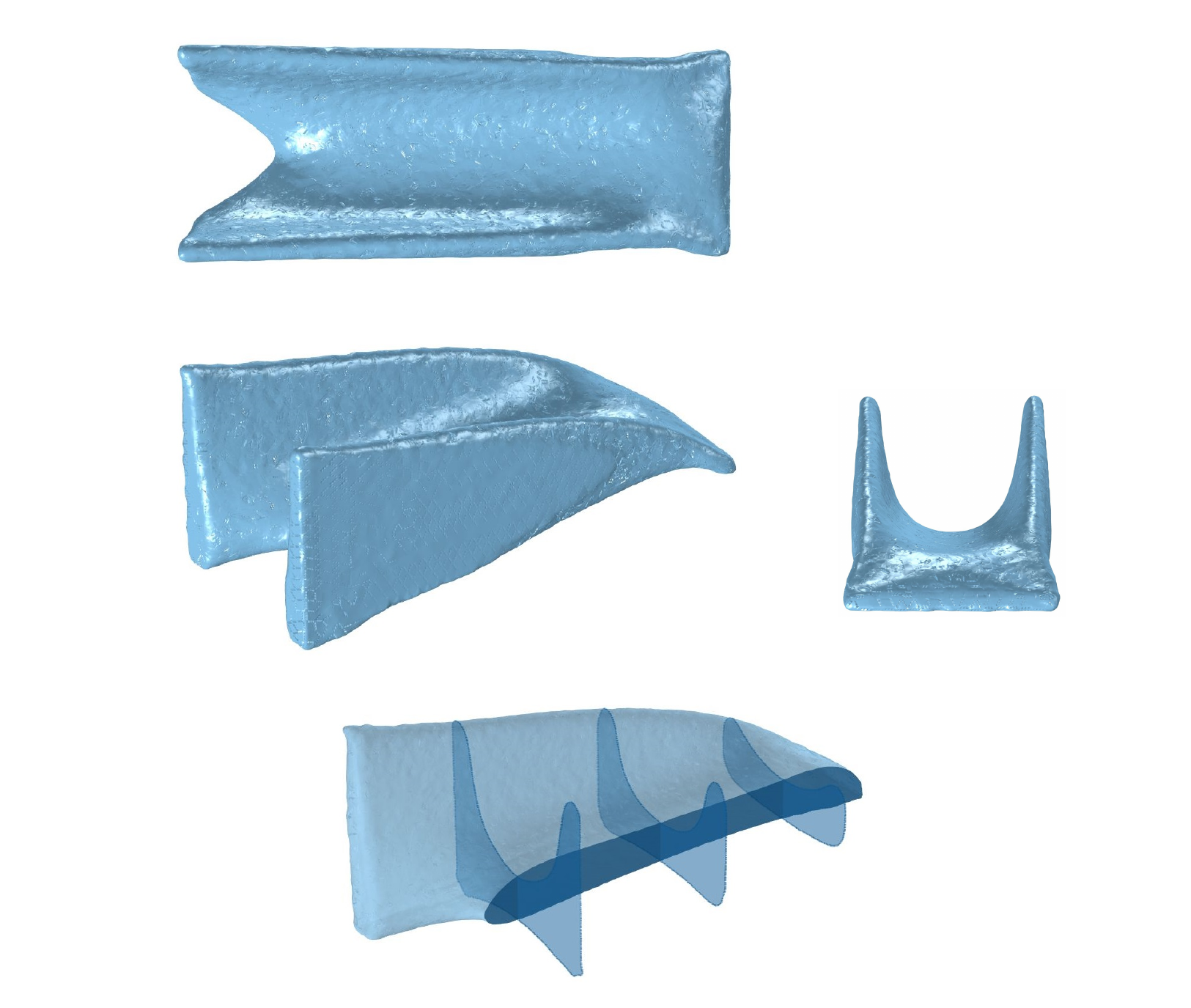
	\caption{The baseline optimized sheared beam design with the $0.25$ volume fraction setting by the conventional topology optimization method (compliance: $0.48$). (a) Top view. (b) Isometric view. (c) Right view. (d) Cross-sectional view.}
\label{fig:3dcanti_simp}
\end{figure}

The first example is a cuboid design domain with a fixed (in all three degrees of freedom) face and a sheared edge load. Its detailed design domain and boundary condition settings are presented in Figure~\ref{fig:3d_bc}(a). With the volume fraction set as $0.25$, Figure~\ref{fig:3dcanti_simp} presents the baseline design optimized by the conventional topology optimization method. A U-shaped beam is generated with variable-shaped cross sections. From the cross-sectional view, it is observed that the optimized beam structure is simply-connected (i.e., no enclosed cavities). The thickness of the structure is not constant. Its optimized compliance value is $0.48$.

\begin{figure}
    \centering
    \scriptsize
	\def\svgwidth{0.85\linewidth}
	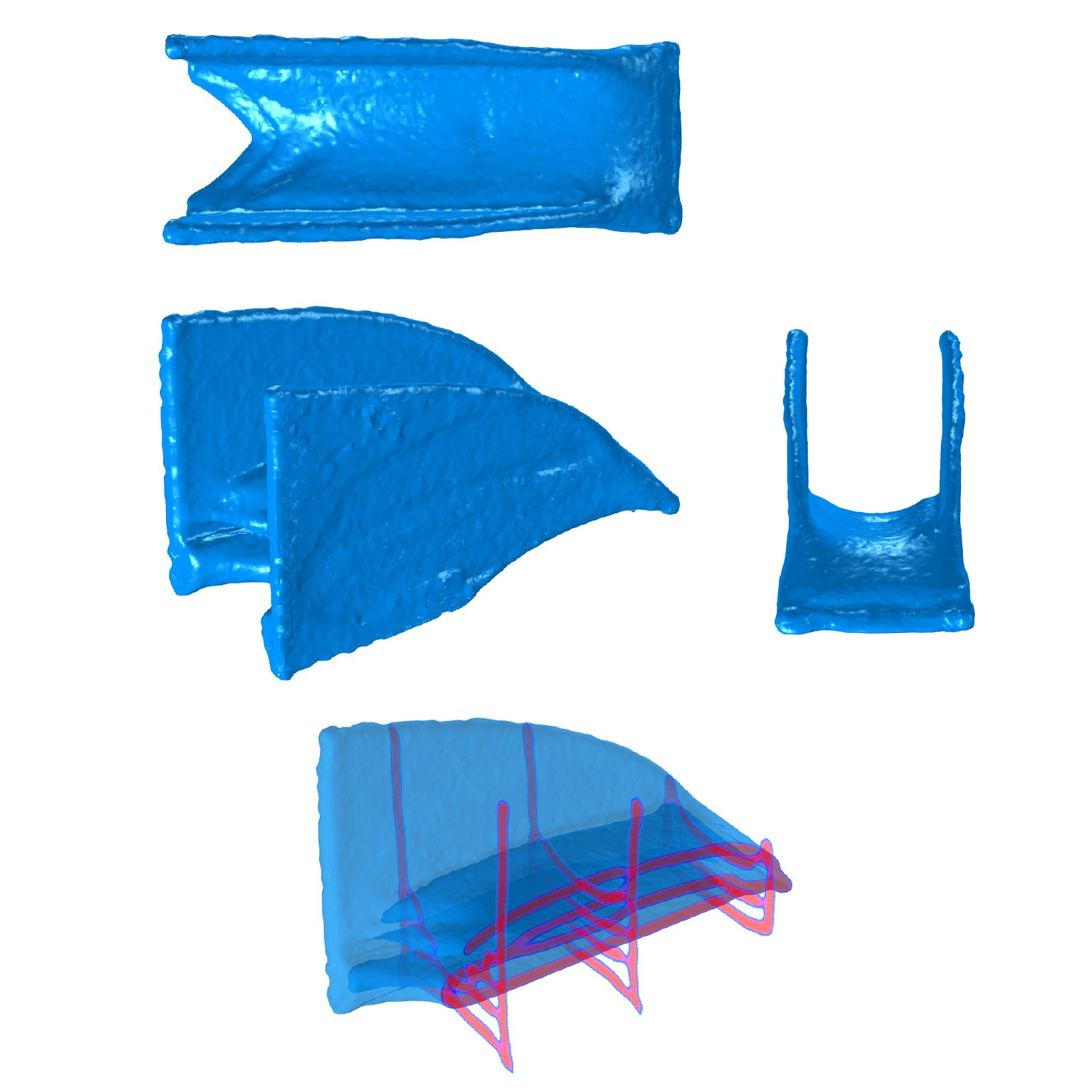
	\caption{The optimized sheared beam thin-walled design with the $0.25$ volume fraction setting (compliance: $0.51$). (a) Top view. (b) Isometric view. (c) Right view. (d) Cross-sectional view.}
\label{fig:3dcanti_25}
\end{figure}

Figure~\ref{fig:3dcanti_25} presents the optimized thin-walled design using the proposed uniform feature size control method. The target volume fraction is $0.25$, the same as that of the baseline design. The resulting thin-walled design is multiply-connected that contains several branched walls and enclosed cavities. Its topology and shape are different from the baseline design. Due to the additional geometric restriction applied to the thin-walled design, its optimized compliance value is $0.51$, which is slightly inferior to that of the baseline design.

\begin{figure}
    \centering
    \scriptsize
	\def\svgwidth{0.85\linewidth}
	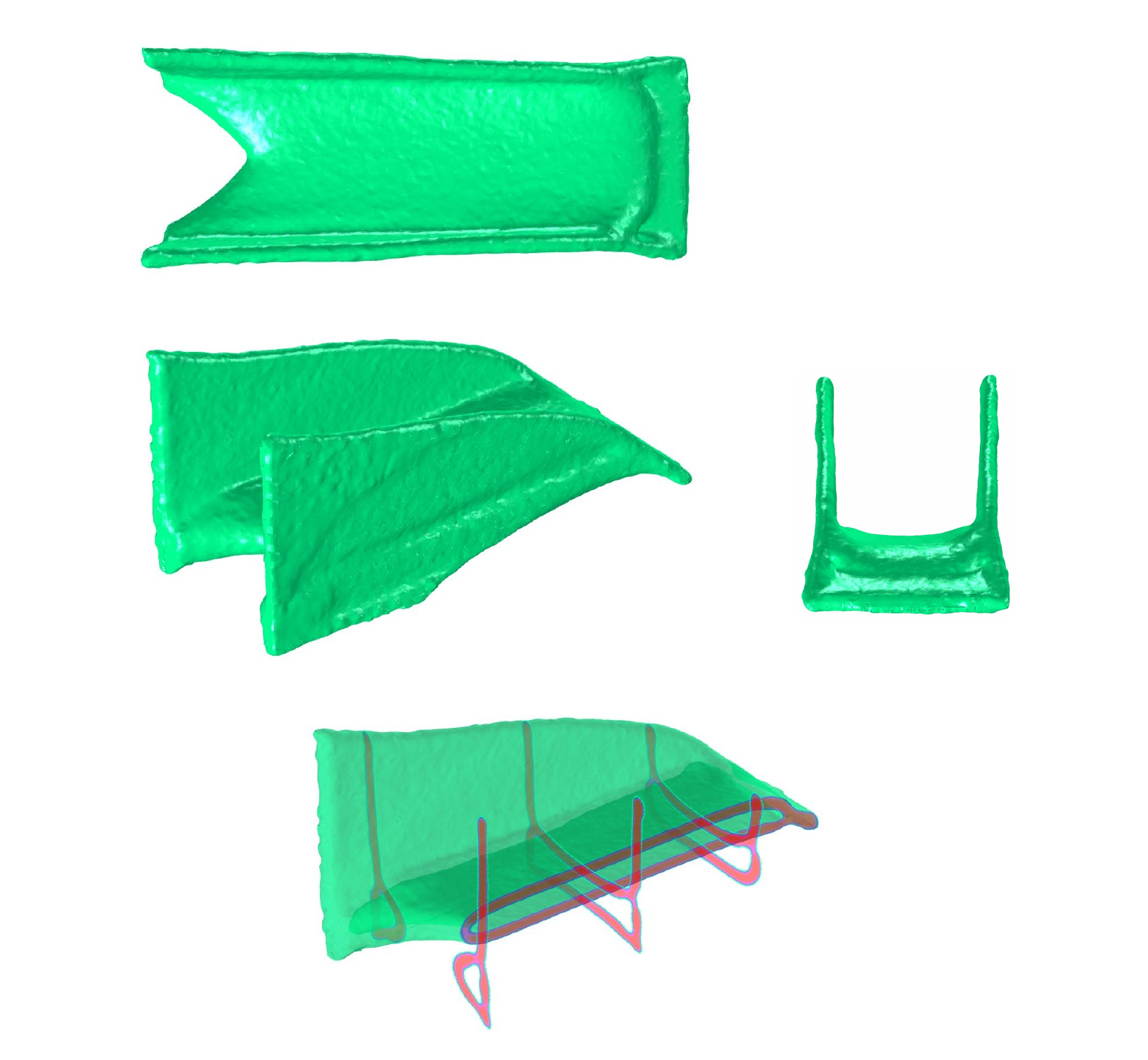
	\caption{The optimized sheared beam thin-walled design with the $0.15$ volume fraction setting. (a) Top view. (b) Isometric view. (c) Right view. (d) Cross-sectional view.}
\label{fig:3dcanti_15}
\end{figure}

\begin{figure}
    \centering
    \scriptsize
	\def\svgwidth{.95\linewidth}
	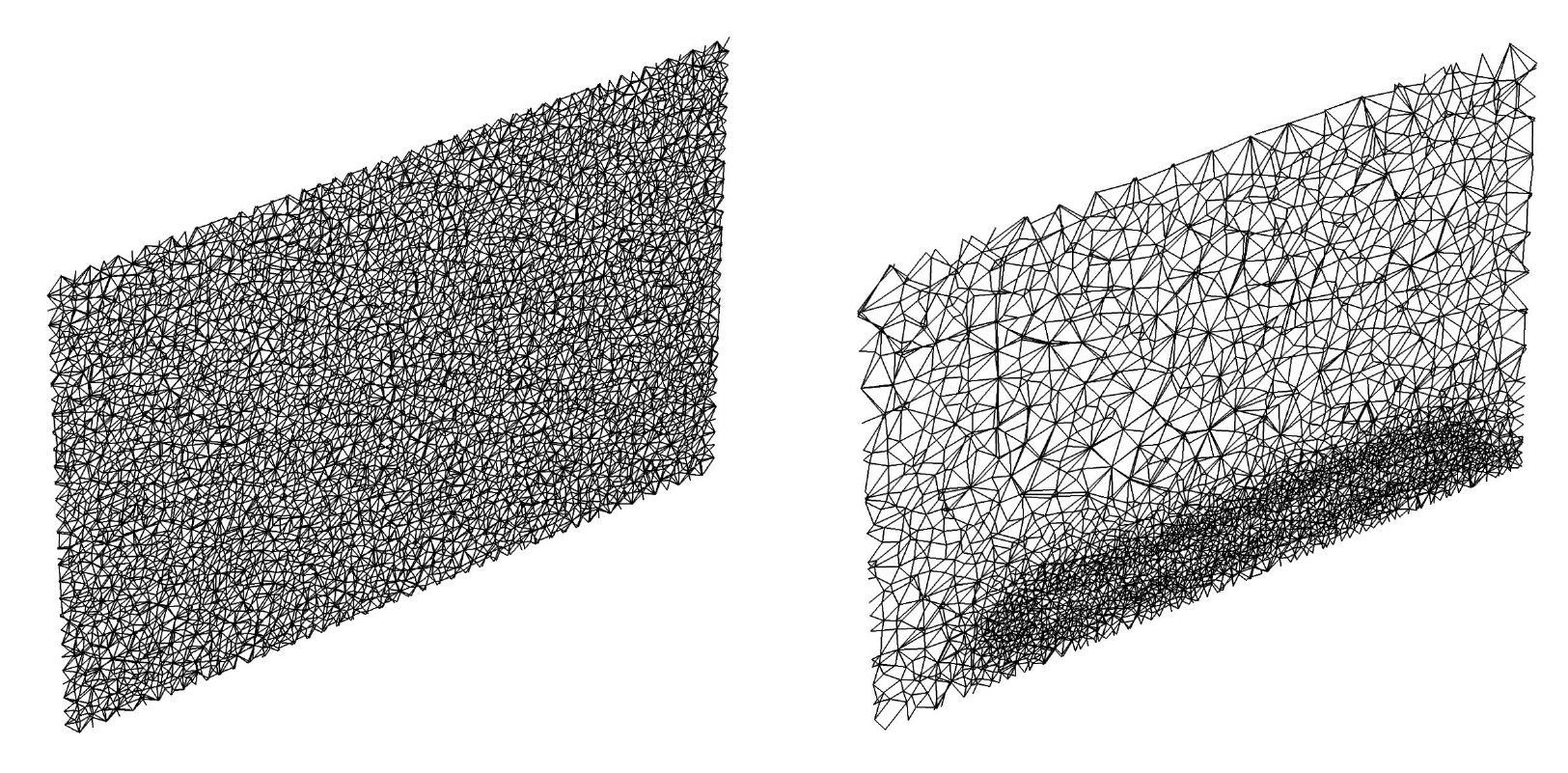
	\caption{Adaptive mesh configuration plots with cross-sectional views at the mid-plane of $y=0.3$. (a) Initial uniform mesh at the start of optimization. (b) Adapted mesh at the end of optimization. (Number of elements: (a) 434703, (b) 330972.)}
\label{fig:3dcanti_15_mesh}
\end{figure}

With the smaller volume fraction of $0.15$, Figure~\ref{fig:3dcanti_15} presents another optimized thin-walled design of the sheared beam problem. Due to the smaller amount of materials allowed, the number of branched walls is reduced. The computational cost saving due to the adaptive mesh is greater for smaller volume fraction problems. To demonstrate the potential number of elements saved with the adaptive mesh scheme, Figure~\ref{fig:3dcanti_15_mesh} shows the domain discretization at the mid-plane of $y=0.3$ for (a) the initial uniform mesh at the start of optimization, and (b) the adapted mesh at the end of optimization. Their element counts are $434703$ and $330972$, respectively. As seen in Figure~\ref{fig:3dcanti_15_mesh}(b), regions with high densities ({\em i.e.}, materials) have finer mesh than those with low densities ({\em i.e.}, voids). The smallest element size at the end of optimization is smaller than that of the initial uniform mesh size. If the design domain were to be meshed with such fine elements uniformly and kept fixed during the course of optimization, the total number of elements required would have been several times more (depending on adaptive mesh settings and the target volume fraction).

\subsection{Twisted ball}
\begin{figure}
    \centering
    \scriptsize
	\def\svgwidth{0.8\linewidth}
	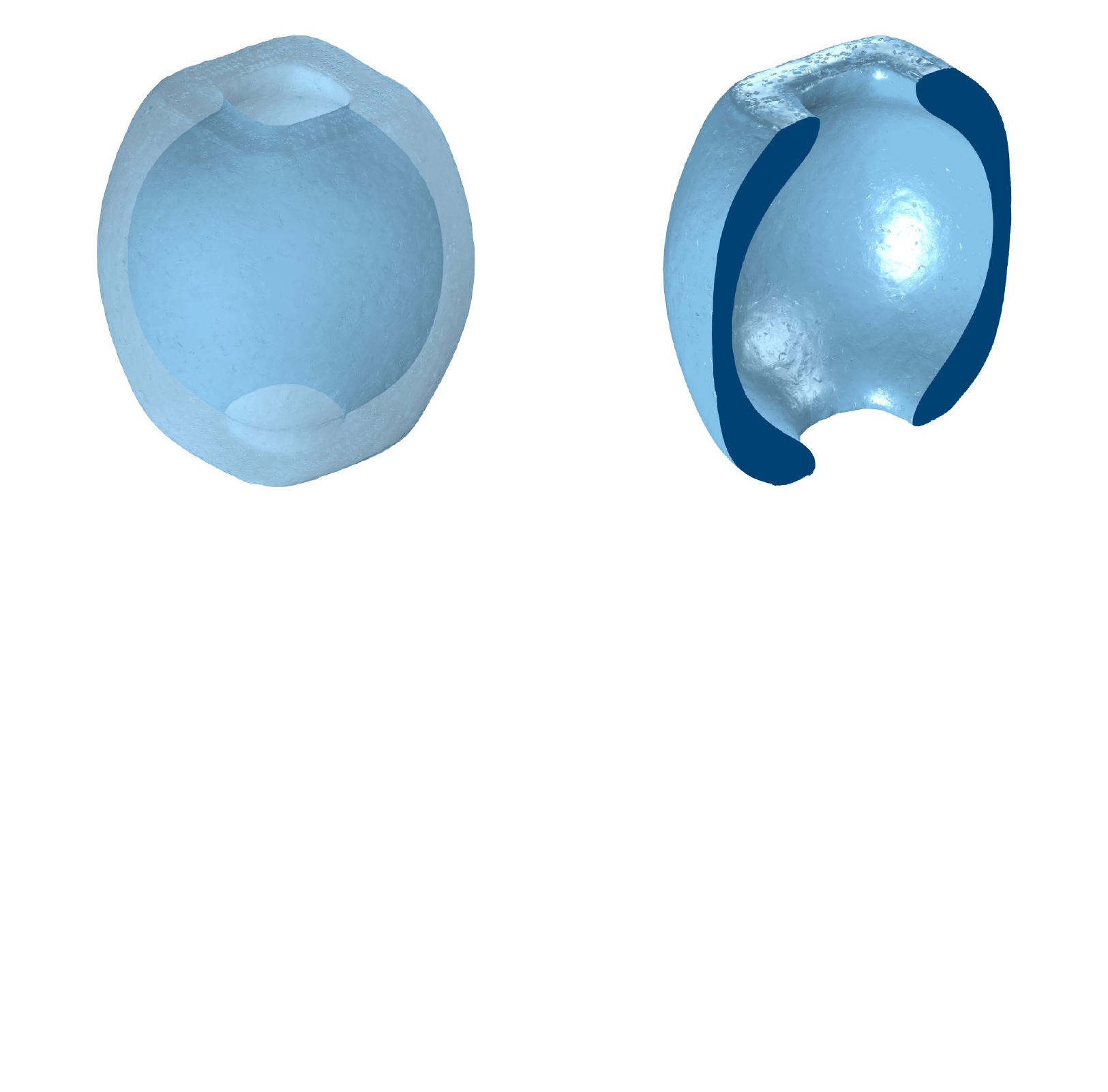
	\caption{The optimized twisted ball designs with the $0.25$ volume fraction setting. (a) The baseline design by the conventional topology optimization method (compliance objective: $0.71$). (b) The thin-walled design (compliance objective: $0.78$).}
\label{fig:3dtorque_35}
\end{figure}

The second example is a boxed design domain with a twist load on the top and a fixed (in all three degrees of freedom) face in the bottom center. Its detailed design domain and boundary condition settings are presented in Figure~\ref{fig:3d_bc}(b). With the volume fraction set as $0.25$, Figure~\ref{fig:3dtorque_35}(a) presents the baseline design optimized by the conventional topology optimization method. A variable-thickness egg shell structure is generated. Its optimized compliance value is $0.71$. With the same volume fraction setting as the baseline design, Figure~\ref{fig:3dtorque_35}(b) presents the thin-walled design with the proposed uniform thickness constraint. A double shell design is obtained due to the enforced thin thickness. As expected, the optimized compliance value is $0.78$, which is slightly worse than the baseline design due to the added thin wall geometric constraint.

\begin{figure}
    \centering
    \scriptsize
	\def\svgwidth{0.8\linewidth}
	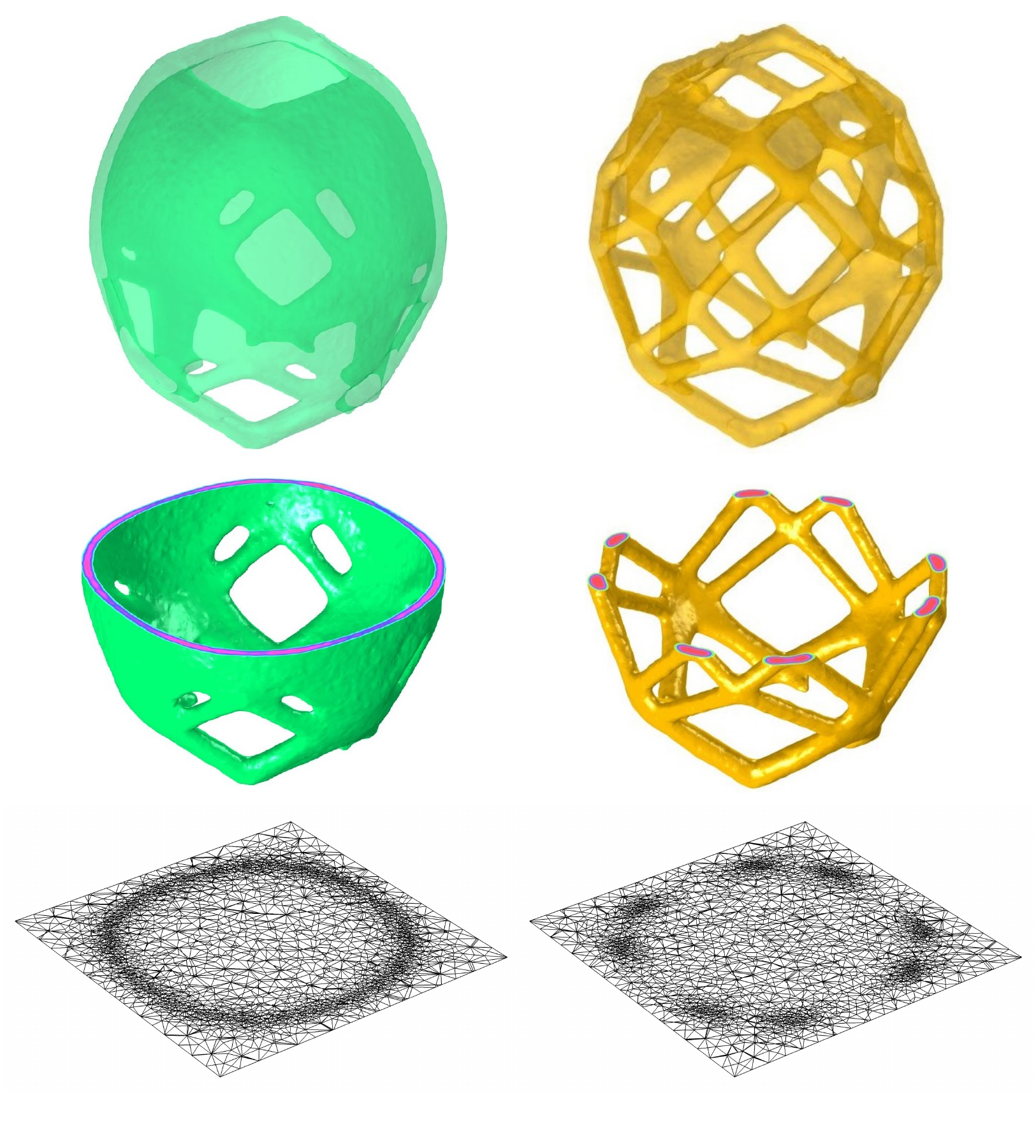
	\caption{The optimized twisted ball thin-walled designs with the (a) $0.08$ and (b) $0.05$ volume fraction settings. The adaptive mesh configuration plots show the cross-sectional views at the $z=0.5$ plane.}
\label{fig:3dtorque_8_5}
\end{figure}

With smaller volume fraction of $0.08$ and $0.05$, Figure~\ref{fig:3dtorque_8_5}(a) and Figure~\ref{fig:3dtorque_8_5}(b) present the corresponding optimized designs. In these cases, the maximum feature size has almost no effect. Because the allowable material amount is less than that of a single shell design with a thickness of the prescribed minimum feature size. Mitchell-like structures are created by optimally removing materials on the single enclosed egg shell. The two smaller volume fraction results are consistent with similar problems previously reported~\citep{aage2015topology}. However, it is noted that such 3D thin-walled designs with a fully meshed solid domain and low volume fraction settings can be computationally challenging to obtain without the use of multi-core paralleled high performance computing resources. This paper utilized the adaptive mesh scheme that realized the high-resolution thin-walled designs using only a standard desktop PC (CPU: Xeon E3-1241 v3 3.5GHz; RAM: 16 GB).

\subsection{Multi-direction loaded cube}
Finally, a more complicated multi-directional loading condition problem is presented. The boundary loads include a vertical force and a shear force, which are simultaneously applied to a cubic design domain. One of four bottom faces is fixed in all three degrees of freedom while the rest are fixed only in the $z$ direction. Its detailed design domain and boundary condition settings are presented in Figure~\ref{fig:3d_bc}(c). The target volume fraction is $0.15$.

\begin{figure}
    \centering
    \scriptsize
	\def\svgwidth{0.8\linewidth}
	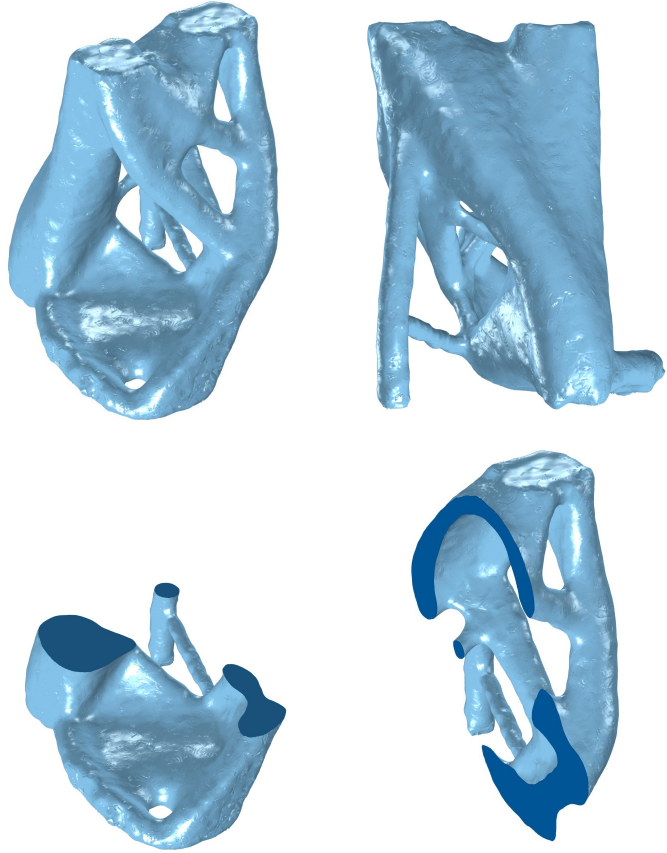
	\caption{The baseline optimized multi-direction loaded cube design by the conventional topology optimization method (compliance objective: $0.714$)}
\label{fig:3dmulti}
\end{figure}

\begin{figure}
    \centering
    \scriptsize
	\def\svgwidth{0.8\linewidth}
	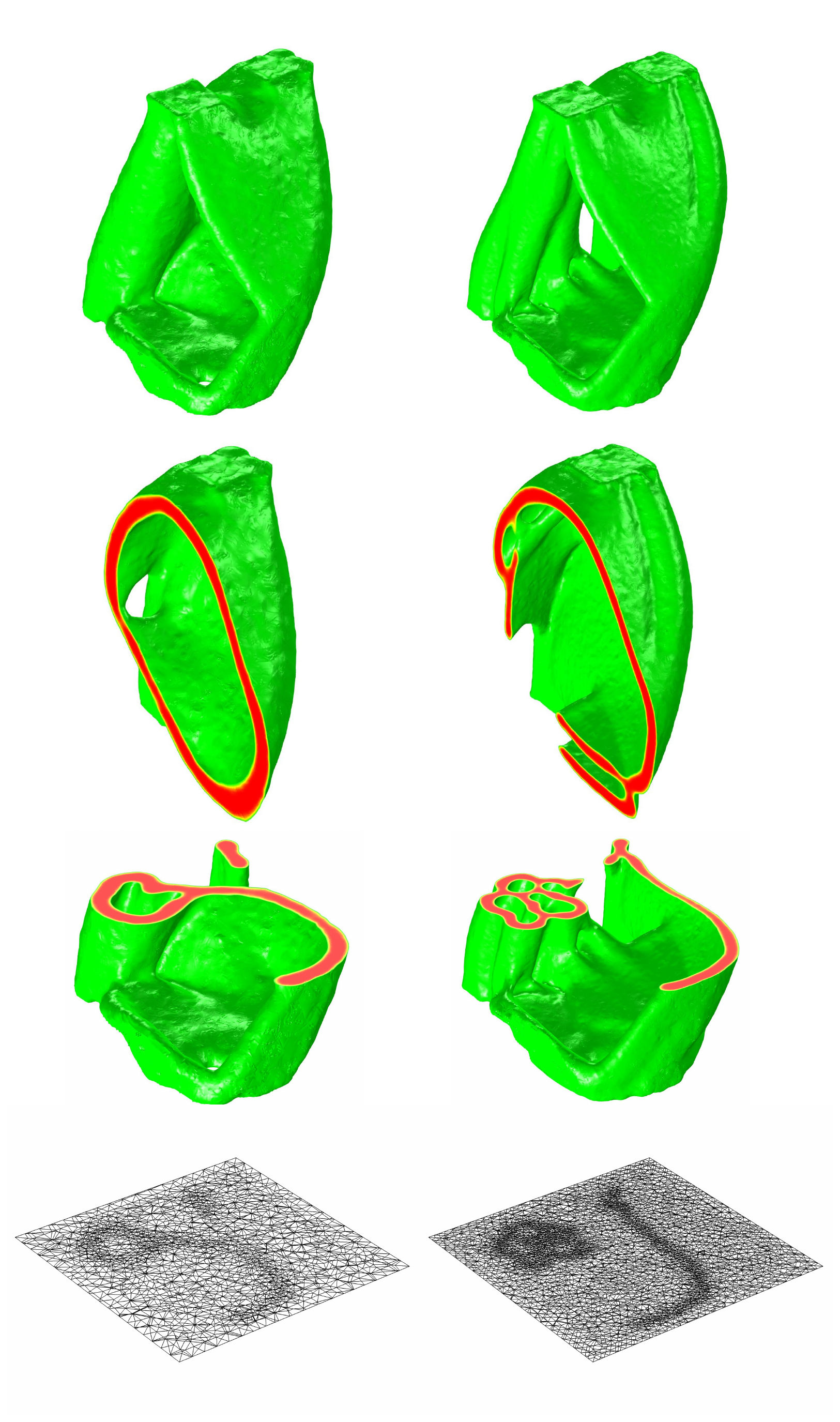
	\caption{The optimized thin-walled designs for the multi-direction loaded cube example with different prescribed thicknesses. (a) Thicker, (b) thinner. Compliance objectives: (a) $0.722$, (b) $0.754$. The adaptive mesh configuration plots show the cross-sectional views at the $z=0.3$ plane.}
\label{fig:3dmulti_shell}
\end{figure}

The optimized conventional topology optimization design is presented in Figure~\ref{fig:3dmulti}, which mostly consists of variable cross-section bars with many small openings. Two thin-walled designs are presented in Figure~\ref{fig:3dmulti_shell} with different prescribed thicknesses. The thin-walled designs, on the other hand, has more continuous and enclosed wall features with fewer yet larger openings. It is noted that branches, curved walls, and hole cut-outs all naturally appeared in thin-walled designs. As seen in Figure~\ref{fig:3dmulti_shell}(b), as the prescribed wall thickness becomes thinner while the total volume is kept constant, reinforcement walls appeared in some local regions. As in all previous examples, the optimized thin-walled designs have slightly worse structural compliance performance than that of the conventional design. The optimized compliance objective for the baseline design is $0.714$. It is observed that as the wall thickness gets thinner, the structural compliance performance of thin-walled designs becomes worse. Their optimized compliance objectives are $0.722$ and $0.754$.

As the thickness becomes thinner, the design domain requires finer discretization, which takes longer computational time. The number of elements at the beginning of optimization (i.e., uniform) and at the end of optimization (i.e. adaptive) are reported in Table~\ref{tab:multi} along with the run time [hours] of 100 optimization iterations (CPU: Xeon E5-1680 v4 3.4 GHz; RAM: 128 GB).
\begin{table}[t]
	\caption{Computational performance summary of the multi-direction loaded cube example}
	\begin{center}
	\label{tab:multi}
	\begin{tabular}{c l l l l}
		\hline
		& \#elem (init) & \#elem (end) & Run time\\
		\hline
		Baseline (Fig~\ref{fig:3dmulti}) &$240453$ &$190713$ &1.2h\\
		Shell (Fig~\ref{fig:3dmulti_shell}(a)) &$240453$ &$185984$ &1.3h\\
		Shell (Fig~\ref{fig:3dmulti_shell}(b)) &$950124$ &$797230$ &7.0h\\
		\hline
	\end{tabular}
	\end{center}
\end{table}

%% file: figures/3d_bc.pdf_tex
\begingroup%
  \makeatletter%
  \providecommand\color[2][]{%
    \errmessage{(Inkscape) Color is used for the text in Inkscape, but the package 'color.sty' is not loaded}%
    \renewcommand\color[2][]{}%
  }%
  \providecommand\transparent[1]{%
    \errmessage{(Inkscape) Transparency is used (non-zero) for the text in Inkscape, but the package 'transparent.sty' is not loaded}%
    \renewcommand\transparent[1]{}%
  }%
  \providecommand\rotatebox[2]{#2}%
  \newcommand*\fsize{\dimexpr\f@size pt\relax}%
  \newcommand*\lineheight[1]{\fontsize{\fsize}{#1\fsize}\selectfont}%
  \ifx\svgwidth\undefined%
    \setlength{\unitlength}{468bp}%
    \ifx\svgscale\undefined%
      \relax%
    \else%
      \setlength{\unitlength}{\unitlength * \real{\svgscale}}%
    \fi%
  \else%
    \setlength{\unitlength}{\svgwidth}%
  \fi%
  \global\let\svgwidth\undefined%
  \global\let\svgscale\undefined%
  \makeatother%
  \begin{picture}(1,0.92307692)%
    \lineheight{1}%
    \setlength\tabcolsep{0pt}%
    \put(-0.11904762,0.86996349){\color[rgb]{0,0,0}\transparent{0.2}\makebox(0,0)[lt]{\begin{minipage}{1.29349825\unitlength}\raggedright \end{minipage}}}%
    \put(0,0){\includegraphics[width=\unitlength,page=1]{3d_bc.pdf}}%
    \put(0.10060768,0.10802357){\color[rgb]{0,0,0}\makebox(0,0)[lt]{\lineheight{1.25}\smash{\begin{tabular}[t]{l}$x$\end{tabular}}}}%
    \put(0.02579924,0.10960398){\color[rgb]{0,0,0}\makebox(0,0)[lt]{\lineheight{1.25}\smash{\begin{tabular}[t]{l}$y$\end{tabular}}}}%
    \put(0.05830606,0.1397918){\color[rgb]{0,0,0}\makebox(0,0)[lt]{\lineheight{1.25}\smash{\begin{tabular}[t]{l}$z$\end{tabular}}}}%
    \put(0.31623853,0.35336085){\color[rgb]{0,0,0}\makebox(0,0)[lt]{\lineheight{1.25}\smash{\begin{tabular}[t]{l}$F$\end{tabular}}}}%
    \put(0,0){\includegraphics[width=\unitlength,page=2]{3d_bc.pdf}}%
    \put(0.1400235,0.25476495){\color[rgb]{0,0,0}\makebox(0,0)[lt]{\lineheight{1.25}\smash{\begin{tabular}[t]{l}$1$\end{tabular}}}}%
    \put(0.20950906,0.11198853){\color[rgb]{0,0,0}\rotatebox{-38.888414}{\makebox(0,0)[lt]{\lineheight{1.25}\smash{\begin{tabular}[t]{l}$1$\end{tabular}}}}}%
    \put(0.3951978,0.08661682){\color[rgb]{0,0,0}\rotatebox{24.089161}{\makebox(0,0)[lt]{\lineheight{1.25}\smash{\begin{tabular}[t]{l}$1$\end{tabular}}}}}%
    \put(0.30379931,0.02835624){\color[rgb]{0,0,0}\makebox(0,0)[lt]{\lineheight{1.25}\smash{\begin{tabular}[t]{l}(b)\end{tabular}}}}%
    \put(0.71758557,0.02181631){\color[rgb]{0,0,0}\makebox(0,0)[lt]{\lineheight{1.25}\smash{\begin{tabular}[t]{l}(c)\end{tabular}}}}%
    \put(0,0){\includegraphics[width=\unitlength,page=3]{3d_bc.pdf}}%
    \put(0.56310104,0.25156017){\color[rgb]{0,0,0}\makebox(0,0)[lt]{\lineheight{1.25}\smash{\begin{tabular}[t]{l}$1$\end{tabular}}}}%
    \put(0.63258178,0.10878793){\color[rgb]{0,0,0}\rotatebox{-38.888414}{\makebox(0,0)[lt]{\lineheight{1.25}\smash{\begin{tabular}[t]{l}$1$\end{tabular}}}}}%
    \put(0.818271,0.08341928){\color[rgb]{0,0,0}\rotatebox{24.089161}{\makebox(0,0)[lt]{\lineheight{1.25}\smash{\begin{tabular}[t]{l}$1$\end{tabular}}}}}%
    \put(0,0){\includegraphics[width=\unitlength,page=4]{3d_bc.pdf}}%
    \put(0.72394075,0.29191509){\color[rgb]{0,0,0}\makebox(0,0)[lt]{\lineheight{1.25}\smash{\begin{tabular}[t]{l}$2F$\end{tabular}}}}%
    \put(0.72697129,0.38692809){\color[rgb]{0,0,0}\makebox(0,0)[lt]{\lineheight{1.25}\smash{\begin{tabular}[t]{l}$F$\end{tabular}}}}%
    \put(0,0){\includegraphics[width=\unitlength,page=5]{3d_bc.pdf}}%
    \put(0.73506709,0.56302358){\color[rgb]{0,0,0}\makebox(0,0)[lt]{\lineheight{1.25}\smash{\begin{tabular}[t]{l}$F$\end{tabular}}}}%
    \put(0.45416778,0.83208624){\color[rgb]{0,0,0}\rotatebox{19.322152}{\makebox(0,0)[lt]{\lineheight{1.25}\smash{\begin{tabular}[t]{l}$1.6$\end{tabular}}}}}%
    \put(0.67541086,0.86870025){\color[rgb]{0,0,0}\rotatebox{-29.723585}{\makebox(0,0)[lt]{\lineheight{1.25}\smash{\begin{tabular}[t]{l}$0.6$\end{tabular}}}}}%
    \put(0.28022238,0.65126199){\color[rgb]{0,0,0}\makebox(0,0)[lt]{\lineheight{1.25}\smash{\begin{tabular}[t]{l}$1$\end{tabular}}}}%
    \put(0.55185903,0.48063767){\color[rgb]{0,0,0}\makebox(0,0)[lt]{\lineheight{1.25}\smash{\begin{tabular}[t]{l}(a)\end{tabular}}}}%
  \end{picture}%
\endgroup%

%% file: figures/3dcanti_simp.pdf_tex
\begingroup%
  \makeatletter%
  \providecommand\color[2][]{%
    \errmessage{(Inkscape) Color is used for the text in Inkscape, but the package 'color.sty' is not loaded}%
    \renewcommand\color[2][]{}%
  }%
  \providecommand\transparent[1]{%
    \errmessage{(Inkscape) Transparency is used (non-zero) for the text in Inkscape, but the package 'transparent.sty' is not loaded}%
    \renewcommand\transparent[1]{}%
  }%
  \providecommand\rotatebox[2]{#2}%
  \newcommand*\fsize{\dimexpr\f@size pt\relax}%
  \newcommand*\lineheight[1]{\fontsize{\fsize}{#1\fsize}\selectfont}%
  \ifx\svgwidth\undefined%
    \setlength{\unitlength}{468bp}%
    \ifx\svgscale\undefined%
      \relax%
    \else%
      \setlength{\unitlength}{\unitlength * \real{\svgscale}}%
    \fi%
  \else%
    \setlength{\unitlength}{\svgwidth}%
  \fi%
  \global\let\svgwidth\undefined%
  \global\let\svgscale\undefined%
  \makeatother%
  \begin{picture}(1,0.84615385)%
    \lineheight{1}%
    \setlength\tabcolsep{0pt}%
    \put(-0.11904762,0.86996344){\color[rgb]{0,0,0}\makebox(0,0)[lt]{\begin{minipage}{1.29349825\unitlength}\raggedright \end{minipage}}}%
    \put(0,0){\includegraphics[width=\unitlength,page=1]{3dcanti_simp.pdf}}%
    \put(0.37235026,0.59121901){\color[rgb]{0,0,0}\makebox(0,0)[lt]{\lineheight{1.25}\smash{\begin{tabular}[t]{l}(a)\end{tabular}}}}%
    \put(0.43431053,0.34345425){\color[rgb]{0,0,0}\makebox(0,0)[lt]{\lineheight{1.25}\smash{\begin{tabular}[t]{l}(b)\end{tabular}}}}%
    \put(0.78603441,0.29668602){\color[rgb]{0,0,0}\makebox(0,0)[lt]{\lineheight{1.25}\smash{\begin{tabular}[t]{l}(c)\end{tabular}}}}%
    \put(0.54365907,0.03906432){\color[rgb]{0,0,0}\makebox(0,0)[lt]{\lineheight{1.25}\smash{\begin{tabular}[t]{l}(d)\end{tabular}}}}%
  \end{picture}%
\endgroup%

%% file: figures/3dcanti_25.pdf_tex
\begingroup%
  \makeatletter%
  \providecommand\color[2][]{%
    \errmessage{(Inkscape) Color is used for the text in Inkscape, but the package 'color.sty' is not loaded}%
    \renewcommand\color[2][]{}%
  }%
  \providecommand\transparent[1]{%
    \errmessage{(Inkscape) Transparency is used (non-zero) for the text in Inkscape, but the package 'transparent.sty' is not loaded}%
    \renewcommand\transparent[1]{}%
  }%
  \providecommand\rotatebox[2]{#2}%
  \newcommand*\fsize{\dimexpr\f@size pt\relax}%
  \newcommand*\lineheight[1]{\fontsize{\fsize}{#1\fsize}\selectfont}%
  \ifx\svgwidth\undefined%
    \setlength{\unitlength}{468bp}%
    \ifx\svgscale\undefined%
      \relax%
    \else%
      \setlength{\unitlength}{\unitlength * \real{\svgscale}}%
    \fi%
  \else%
    \setlength{\unitlength}{\svgwidth}%
  \fi%
  \global\let\svgwidth\undefined%
  \global\let\svgscale\undefined%
  \makeatother%
  \begin{picture}(1,1)%
    \lineheight{1}%
    \setlength\tabcolsep{0pt}%
    \put(-0.11904762,0.86996337){\color[rgb]{0,0,0}\makebox(0,0)[lt]{\begin{minipage}{1.29349825\unitlength}\raggedright \end{minipage}}}%
    \put(0,0){\includegraphics[width=\unitlength,page=1]{3dcanti_25.pdf}}%
    \put(0.38929165,0.74094418){\color[rgb]{0,0,0}\makebox(0,0)[lt]{\lineheight{1.25}\smash{\begin{tabular}[t]{l}(a)\end{tabular}}}}%
    \put(0.45628859,0.41167752){\color[rgb]{0,0,0}\makebox(0,0)[lt]{\lineheight{1.25}\smash{\begin{tabular}[t]{l}(b)\end{tabular}}}}%
    \put(0.79336121,0.38093527){\color[rgb]{0,0,0}\makebox(0,0)[lt]{\lineheight{1.25}\smash{\begin{tabular}[t]{l}(c)\end{tabular}}}}%
    \put(0.57754186,0.03723283){\color[rgb]{0,0,0}\makebox(0,0)[lt]{\lineheight{1.25}\smash{\begin{tabular}[t]{l}(d)\end{tabular}}}}%
  \end{picture}%
\endgroup%

%% file: figures/3dcanti_15.pdf_tex
\begingroup%
  \makeatletter%
  \providecommand\color[2][]{%
    \errmessage{(Inkscape) Color is used for the text in Inkscape, but the package 'color.sty' is not loaded}%
    \renewcommand\color[2][]{}%
  }%
  \providecommand\transparent[1]{%
    \errmessage{(Inkscape) Transparency is used (non-zero) for the text in Inkscape, but the package 'transparent.sty' is not loaded}%
    \renewcommand\transparent[1]{}%
  }%
  \providecommand\rotatebox[2]{#2}%
  \newcommand*\fsize{\dimexpr\f@size pt\relax}%
  \newcommand*\lineheight[1]{\fontsize{\fsize}{#1\fsize}\selectfont}%
  \ifx\svgwidth\undefined%
    \setlength{\unitlength}{468bp}%
    \ifx\svgscale\undefined%
      \relax%
    \else%
      \setlength{\unitlength}{\unitlength * \real{\svgscale}}%
    \fi%
  \else%
    \setlength{\unitlength}{\svgwidth}%
  \fi%
  \global\let\svgwidth\undefined%
  \global\let\svgscale\undefined%
  \makeatother%
  \begin{picture}(1,0.92307692)%
    \lineheight{1}%
    \setlength\tabcolsep{0pt}%
    \put(-0.11904762,0.86996349){\color[rgb]{0,0,0}\makebox(0,0)[lt]{\begin{minipage}{1.29349825\unitlength}\raggedright \end{minipage}}}%
    \put(0,0){\includegraphics[width=\unitlength,page=1]{3dcanti_15.pdf}}%
    \put(0.33017449,0.65835284){\color[rgb]{0,0,0}\makebox(0,0)[lt]{\lineheight{1.25}\smash{\begin{tabular}[t]{l}(a)\end{tabular}}}}%
    \put(0.38892959,0.36205312){\color[rgb]{0,0,0}\makebox(0,0)[lt]{\lineheight{1.25}\smash{\begin{tabular}[t]{l}(b)\end{tabular}}}}%
    \put(0.78048866,0.35237299){\color[rgb]{0,0,0}\makebox(0,0)[lt]{\lineheight{1.25}\smash{\begin{tabular}[t]{l}(c)\end{tabular}}}}%
    \put(0.50743567,0.04667455){\color[rgb]{0,0,0}\makebox(0,0)[lt]{\lineheight{1.25}\smash{\begin{tabular}[t]{l}(d)\end{tabular}}}}%
  \end{picture}%
\endgroup%

%% file: figures/3dcanti_15_mesh.pdf_tex
\begingroup%
  \makeatletter%
  \providecommand\color[2][]{%
    \errmessage{(Inkscape) Color is used for the text in Inkscape, but the package 'color.sty' is not loaded}%
    \renewcommand\color[2][]{}%
  }%
  \providecommand\transparent[1]{%
    \errmessage{(Inkscape) Transparency is used (non-zero) for the text in Inkscape, but the package 'transparent.sty' is not loaded}%
    \renewcommand\transparent[1]{}%
  }%
  \providecommand\rotatebox[2]{#2}%
  \newcommand*\fsize{\dimexpr\f@size pt\relax}%
  \newcommand*\lineheight[1]{\fontsize{\fsize}{#1\fsize}\selectfont}%
  \ifx\svgwidth\undefined%
    \setlength{\unitlength}{468bp}%
    \ifx\svgscale\undefined%
      \relax%
    \else%
      \setlength{\unitlength}{\unitlength * \real{\svgscale}}%
    \fi%
  \else%
    \setlength{\unitlength}{\svgwidth}%
  \fi%
  \global\let\svgwidth\undefined%
  \global\let\svgscale\undefined%
  \makeatother%
  \begin{picture}(1,0.4923077)%
    \lineheight{1}%
    \setlength\tabcolsep{0pt}%
    \put(-0.11904762,0.86996346){\color[rgb]{0,0,0}\makebox(0,0)[lt]{\begin{minipage}{1.29349828\unitlength}\raggedright \end{minipage}}}%
    \put(-0.45260311,-0.39529769){\color[rgb]{0,0,0}\makebox(0,0)[lt]{\lineheight{1.25}\smash{\begin{tabular}[t]{l}(b)\\\\\\\end{tabular}}}}%
    \put(0,0){\includegraphics[width=\unitlength,page=1]{3dcanti_15_mesh.pdf}}%
    \put(0.26661993,0.03063105){\color[rgb]{0,0,0}\makebox(0,0)[lt]{\lineheight{1.25}\smash{\begin{tabular}[t]{l}(a)\\\\\\\end{tabular}}}}%
    \put(0.78209441,0.02968503){\color[rgb]{0,0,0}\makebox(0,0)[lt]{\lineheight{1.25}\smash{\begin{tabular}[t]{l}(b)\\\\\\\end{tabular}}}}%
  \end{picture}%
\endgroup%

%% file: figures/3dtorque_35.pdf_tex
\begingroup%
  \makeatletter%
  \providecommand\color[2][]{%
    \errmessage{(Inkscape) Color is used for the text in Inkscape, but the package 'color.sty' is not loaded}%
    \renewcommand\color[2][]{}%
  }%
  \providecommand\transparent[1]{%
    \errmessage{(Inkscape) Transparency is used (non-zero) for the text in Inkscape, but the package 'transparent.sty' is not loaded}%
    \renewcommand\transparent[1]{}%
  }%
  \providecommand\rotatebox[2]{#2}%
  \newcommand*\fsize{\dimexpr\f@size pt\relax}%
  \newcommand*\lineheight[1]{\fontsize{\fsize}{#1\fsize}\selectfont}%
  \ifx\svgwidth\undefined%
    \setlength{\unitlength}{468bp}%
    \ifx\svgscale\undefined%
      \relax%
    \else%
      \setlength{\unitlength}{\unitlength * \real{\svgscale}}%
    \fi%
  \else%
    \setlength{\unitlength}{\svgwidth}%
  \fi%
  \global\let\svgwidth\undefined%
  \global\let\svgscale\undefined%
  \makeatother%
  \begin{picture}(1,0.9846154)%
    \lineheight{1}%
    \setlength\tabcolsep{0pt}%
    \put(-0.11904762,0.8699635){\color[rgb]{0,0,0}\makebox(0,0)[lt]{\begin{minipage}{1.29349827\unitlength}\raggedright \end{minipage}}}%
    \put(0.49735711,0.49618103){\color[rgb]{0,0,0}\makebox(0,0)[lt]{\lineheight{1.25}\smash{\begin{tabular}[t]{l}(a)\\\\\\\end{tabular}}}}%
    \put(0,0){\includegraphics[width=\unitlength,page=1]{3dtorque_35.pdf}}%
    \put(0.4954317,0.02232888){\color[rgb]{0,0,0}\makebox(0,0)[lt]{\lineheight{1.25}\smash{\begin{tabular}[t]{l}(b)\\\\\\\end{tabular}}}}%
    \put(0,0){\includegraphics[width=\unitlength,page=2]{3dtorque_35.pdf}}%
  \end{picture}%
\endgroup%

%% file: figures/3dtorque_8_5.pdf_tex
\begingroup%
  \makeatletter%
  \providecommand\color[2][]{%
    \errmessage{(Inkscape) Color is used for the text in Inkscape, but the package 'color.sty' is not loaded}%
    \renewcommand\color[2][]{}%
  }%
  \providecommand\transparent[1]{%
    \errmessage{(Inkscape) Transparency is used (non-zero) for the text in Inkscape, but the package 'transparent.sty' is not loaded}%
    \renewcommand\transparent[1]{}%
  }%
  \providecommand\rotatebox[2]{#2}%
  \newcommand*\fsize{\dimexpr\f@size pt\relax}%
  \newcommand*\lineheight[1]{\fontsize{\fsize}{#1\fsize}\selectfont}%
  \ifx\svgwidth\undefined%
    \setlength{\unitlength}{468bp}%
    \ifx\svgscale\undefined%
      \relax%
    \else%
      \setlength{\unitlength}{\unitlength * \real{\svgscale}}%
    \fi%
  \else%
    \setlength{\unitlength}{\svgwidth}%
  \fi%
  \global\let\svgwidth\undefined%
  \global\let\svgscale\undefined%
  \makeatother%
  \begin{picture}(1,1.10769228)%
    \lineheight{1}%
    \setlength\tabcolsep{0pt}%
    \put(-0.11904762,0.86996346){\color[rgb]{0,0,0}\makebox(0,0)[lt]{\begin{minipage}{1.29349823\unitlength}\raggedright \end{minipage}}}%
    \put(0,0){\includegraphics[width=\unitlength,page=1]{3dtorque_8_5.pdf}}%
    \put(0.2474834,0.0229967){\color[rgb]{0,0,0}\makebox(0,0)[lt]{\lineheight{1.25}\smash{\begin{tabular}[t]{l}(a)\\\\\\\end{tabular}}}}%
    \put(0.7469323,0.02205067){\color[rgb]{0,0,0}\makebox(0,0)[lt]{\lineheight{1.25}\smash{\begin{tabular}[t]{l}(b)\\\\\\\end{tabular}}}}%
  \end{picture}%
\endgroup%

%% file: figures/3dmulti.pdf_tex
\begingroup%
  \makeatletter%
  \providecommand\color[2][]{%
    \errmessage{(Inkscape) Color is used for the text in Inkscape, but the package 'color.sty' is not loaded}%
    \renewcommand\color[2][]{}%
  }%
  \providecommand\transparent[1]{%
    \errmessage{(Inkscape) Transparency is used (non-zero) for the text in Inkscape, but the package 'transparent.sty' is not loaded}%
    \renewcommand\transparent[1]{}%
  }%
  \providecommand\rotatebox[2]{#2}%
  \newcommand*\fsize{\dimexpr\f@size pt\relax}%
  \newcommand*\lineheight[1]{\fontsize{\fsize}{#1\fsize}\selectfont}%
  \ifx\svgwidth\undefined%
    \setlength{\unitlength}{468bp}%
    \ifx\svgscale\undefined%
      \relax%
    \else%
      \setlength{\unitlength}{\unitlength * \real{\svgscale}}%
    \fi%
  \else%
    \setlength{\unitlength}{\svgwidth}%
  \fi%
  \global\let\svgwidth\undefined%
  \global\let\svgscale\undefined%
  \makeatother%
  \begin{picture}(1,1.07692308)%
    \lineheight{1}%
    \setlength\tabcolsep{0pt}%
    \put(-0.11904763,0.86996354){\color[rgb]{0,0,0}\makebox(0,0)[lt]{\begin{minipage}{1.29349832\unitlength}\raggedright \end{minipage}}}%
    \put(0,0){\includegraphics[width=\unitlength,page=1]{3dmulti.pdf}}%
  \end{picture}%
\endgroup%

%% file: figures/3dmulti_shell.pdf_tex
\begingroup%
  \makeatletter%
  \providecommand\color[2][]{%
    \errmessage{(Inkscape) Color is used for the text in Inkscape, but the package 'color.sty' is not loaded}%
    \renewcommand\color[2][]{}%
  }%
  \providecommand\transparent[1]{%
    \errmessage{(Inkscape) Transparency is used (non-zero) for the text in Inkscape, but the package 'transparent.sty' is not loaded}%
    \renewcommand\transparent[1]{}%
  }%
  \providecommand\rotatebox[2]{#2}%
  \newcommand*\fsize{\dimexpr\f@size pt\relax}%
  \newcommand*\lineheight[1]{\fontsize{\fsize}{#1\fsize}\selectfont}%
  \ifx\svgwidth\undefined%
    \setlength{\unitlength}{468bp}%
    \ifx\svgscale\undefined%
      \relax%
    \else%
      \setlength{\unitlength}{\unitlength * \real{\svgscale}}%
    \fi%
  \else%
    \setlength{\unitlength}{\svgwidth}%
  \fi%
  \global\let\svgwidth\undefined%
  \global\let\svgscale\undefined%
  \makeatother%
  \begin{picture}(1,1.69230769)%
    \lineheight{1}%
    \setlength\tabcolsep{0pt}%
    \put(-0.11904763,0.86996348){\color[rgb]{0,0,0}\makebox(0,0)[lt]{\begin{minipage}{1.29349835\unitlength}\raggedright \end{minipage}}}%
    \put(0,0){\includegraphics[width=\unitlength,page=1]{3dmulti_shell.pdf}}%
    \put(0.22968514,0.03555524){\color[rgb]{0,0,0}\makebox(0,0)[lt]{\lineheight{1.25}\smash{\begin{tabular}[t]{l}(a)\\\\\\\end{tabular}}}}%
    \put(0.72913387,0.03460922){\color[rgb]{0,0,0}\makebox(0,0)[lt]{\lineheight{1.25}\smash{\begin{tabular}[t]{l}(b)\\\\\\\end{tabular}}}}%
  \end{picture}%
\endgroup%

%% file: conclusion.tex
This paper presented a Helmholtz PDE-based method that could control both the minimum and the maximum feature sizes in the optimized topologies. By simultaneously enforcing the minimum and the maximum feature sizes, the thin uniform thickness were obtained that generated thin-walled designs in 3D. The general 3D thin-walled designs with the flexibility of branching, curving, and hole cut-outs were demonstrated using a density-based topology optimization method.

As the proposed PDE-based maximum length scale method does not require the neighbor cell information, it enables the efficient computation and is applicable to adaptive meshing implementation. For 3D topology optimization problems with low material-void ratios and thin complex geometric features, the implementation of adaptive meshing significantly saves computational efforts that otherwise would be wasted in large void regions.

3D examples were provided to demonstrate the proposed method in designing general thin-walled structures. The effectiveness of the adaptive meshing were also demonstrated in generating complex 3D thin structures using only standard desktop PCs. While the optimized results have consistently showed the structural performance sacrifice in the thin-walled designs compared to the conventional designs, the resulting thin-walled designs are more practical for many lightweight manufacturing applications, {\em e.g.}, stamping, investment casting and composite manufacturing.

Based on the study presented in this paper, future research can focus on open source and paralleled high performance computing extensions. Additional manufacturing constraints specific to chosen manufacturing processes, {\em e.g.}, stamping, casting and molding, can be added to the proposed formulation, which will generate even more manufacturable designs suitable for these processes.